\documentclass{article}

\pdfoutput=1

\usepackage{graphicx}
\usepackage{natbib}
\usepackage{amsmath,amssymb}
\usepackage{subfig}

\newcommand\Rey{\mbox{\textit{Re}}}  

\newsavebox{\astrutbox}
\sbox{\astrutbox}{\rule[-5pt]{0pt}{20pt}}

\begin{document}

\title{Body-rock or lift-off in flow}

\author{Frank T. Smith\footnote{Department of Mathematics, UCL, Gower Street, London WC1E 6BT, UK.}
and Phillip L. Wilson\footnote{Department of Mathematics \& Statistics, University of Canterbury, Private Bag 4800, Christchurch 8140, New Zealand. phillip.wilson@canterbury.ac.nz}}

\maketitle

\begin{abstract}
Conditions are investigated under which a body lying at rest or rocking on a solid horizontal surface can be removed from the surface by hydrodynamic forces or instead continues rocking. The investigation is motivated by recent observations on Martian dust movement as well as other small- and large-scale applications. The nonlinear theory of fluid-body interaction here has unsteady motion of an inviscid fluid interacting with a moving thin body. Various shapes of body are addressed together with a range of initial conditions. The relevant parameter space is found to be subtle as evolution and shape play substantial roles coupled with scaled mass and gravity effects. Lift-off of the body from the surface generally cannot occur without fluid flow but it can occur either immediately or within a finite time once the fluid flow starts up: parameters for this are found and comparisons are made with Martian observations.
\end{abstract}

\section{Introduction}\label{sec:intro}

The effect or the use of a flow of fluid to remove a body originally stationary or rocking on a fixed solid surface has many applications and is of interest over a range of length and time scales. The force required to do this ‘washing’, ‘clearance’ or ‘erosion’ of the original surface is also of much concern. 

The applications vary from removal of debris, erosion of soil by wind, water or raindrops, sand and pebble movement on beaches, dust loss, dust blowing, leaf-blowers and related geological and industrial phenomena, through cleaning and washing processes such as with fluid knives and cameras, conveyor design, biomedical problems, weather damage, to aircraft take-off, estimation of runway length, sports applications such as ski-jumping, car and cycle racing, and the safety of wind-blown buildings. See for example \citet{godonestanchi11}, \citet{bascom80}, \citet{virmavirtaetal01}, \citet{wittetal99}. In geological settings this process is known as \emph{saltation}.  The typical scales range from the comparatively small sizes of various biomedical settings, dust problems and household/industrial cleaning to the larger sizes of tornados and tsunamis lifting relatively large obstacles \citep{hunt05,mikamietal12}, and related disasters, along with military applications, and even on to planetary size. Moreover, \citet{coxetal12}, \citet{halletal10} and references therein provide observational evidence that shoreline boulders whose weight is of the order of 10 tonnes can be moved by wave motion even in years without exceptional storm activity, and that boulders up to 78 tonnes have been moved to 11m above high water in significant storms, with the role of tsunamis ruled out. A comparison between boulder transport by storm waves and tsunamis is given in \citet{barbanoetal10}. In these geological applications, some of the more impressive feats of wave transport occur with long, slender boulders conforming to the assumptions of the present study. Further, shallow granular avalanches --- such as those occuring in pyroclastic flows and snow avalanches --- are studied in  \citet{johnsongray11}, \citet{grayancey11} where attention is drawn to the effects of stationary granular material within evolving flows, in addition to pile collapse behaviour and segregation rates between constituents, where large and small particles percolate to the base or surface of an avalanche. In the human body, the lift-off due to hydrodynamic forces in the blood flow of thromboses formed near the site of blood-vessel injury, such as that caused by atherosclerosis, is of significant concern as the thrombus or its resulting transported embolus can lead to stroke, myocardial infarction, deep vein thrombosis, and other damaging conditions \citep{ku97}. Hemodynamic lift-off and transport of tumour cells from damaged vascular walls may be one of several rate-limiting factors in cancer metastasis \citep{koumoutsakosetal13}. A dramatic example from human engineering and exploration was the loss of the space shuttle Columbia as a result of its collision with a relatively small piece of insulating foam separated from a fuel tank apparently by the flow past the shuttle \citep{gomezetal03}. We should also mention human experience and pleasure, for example in blowing spilt salt off a table, and watching ``dust devils'' dance across a beach or even Mars.

The issues involved include the need for understanding of the influences of length and time scales, the effects of the incident flow direction and velocity as well as the effects of the detailed obstacle shape, and issues of stall, lift, impact and drag. The roles of the Froude or Richardson, Reynolds and effective Stokes or Womersley numbers in particular need consideration. A related issue is the question of whether or under what circumstances the effects are mostly dominated by momentum considerations or not and to what extent initial values and the variability of the incident flow exert control on the dynamics. Many of the applications mentioned have complex interactions which are of multi-body type, with multiple impacts and rebounds of transported bodies, as well as of fluid-body-interaction type. It may be that considering the dynamics of a single body on its own is not immediately relevant to some of the real-world situations but on the other hand the single-body configuration does create a clear and basic growth point. A hope or hypothesis in principle is that this would develop fully to the multi-body arrangements.

Clearly the criteria for lift-off of the body from the surface are a central issue to be addressed as part of the general picking-up of the body and transporting it. This is so for many applications indeed but includes especially the matters raised in recent work about dust motions, saltation and reptation or splashing of particles on or near the surface of Mars. In particular, the first extraterrestrial measurements of sand transport rates were given in \citet{bridgesetal12a,bridgesetal12b}, and were rather surprisingly estimated to be comparable to those on Earth, in spite of the Martian atmosphere being 100 times less dense than the terrestrial one. These high transport rates occur despite only rare occasions on which the surface wind stresses on Mars are observed to exceed a supposed critical threshold for sand transport: \citet{zimbelman00,haberleetal03}. It has been hypothesised that the relatively low drag and gravity effects could keep sand in motion for longer as it bounces and tumbles across the surface \citet{kok10a,kok10b,kok12}, to which the present study is directly applicable. Equally, we can here address the interesting question raised by \citet{sullivanetal08} as to whether the aerodynamic effects studied herein contribute to the lifting of Martian sands \citep{wang12}.  More general work on dust suspended in the Martian atmospheric boundary layer is also of interest here \citep{tayloretal07}, \citep{davyetal09}. We shall return to discuss the Martian matters in some detail near the end of the present contribution. The phenomena involved in the broad area are dependent on many parameters for sure but are also dynamic, evolutionary and initial-value dependent to one degree or another. The present model which is depicted in figure \ref{fig:config} below and the ensuing working are based on considering theoretically phenomena which are indeed dominated by momentum and pressure forces and hence centre on essentially inviscid unsteady fluid-body interactions of the type studied recently by \citet{smithwilson11}, \citet{hickssmith11}, \citet{smithellis10} in various contexts. These last papers find that due largely to the actions of added mass in the interactions induced between body and fluid interesting real-world phenomena such as touchdown, lift-off, clashing, skimming and rebounds emerge as part of the theory. Relatively few ingredients are necessary but the detailed evolution does tend to play a substantial role.

The fluid is taken here to be Newtonian and incompressible with uniform density $\rho_D$ say, where the subscript $D$ refers to a dimensional quantity. The generally unsteady motion of both the fluid and the immersed body is assumed to be two-dimensional as a starting point for the theory even though it must be accepted that this restriction leaves out the distinct possibility of fluid skirting around a contact point in the third spatial dimension. The representative Reynolds number \Rey\ based on incident flow speed and a typical body length varies from application to application above but the typical \Rey\ value of interest is quite large and so as a first approximation an inviscid separation-free theory is applied, in keeping with the overwhelming nature of the momentum described previously for these fluid-body interactions. Neglect of viscous effects seems acceptable since most of the fluid-body interactions in reality are of the turbulent kind rather than the more sensitive laminar type. Thin bodies are also of interest in their own right because they can lead to fulsome analytical descriptions for many configurations such as in the previous paragraph but further they shed some light on interactions for thicker bodies such as smooth bluff cylinders or spheres wherever thin layers of fluid lie between the body and the supporting solid surface.

The paper itself is structured as follows. Section \ref{sec:fbi} considers the model in detail including particularly the generation of fluid-body interactions, followed by section \ref{sec:smallt} which turns to an analysis for relatively small times and investigates the possibility of an instant lift-off of the body from the solid surface when the oncoming fluid stream is instigated. Numerical studies for order-one times are presented in section \ref{sec:ord1tnoflow} for cases where lift-off either is absent or possibly is delayed, with fluid-flow effects being suppressed. Sections \ref{sec:tord1}, \ref{sec:liftoff} then include numerically and analytically the added influences of fluid flow and the mechanism and criteria for lift-off to take place, respectively. Certain parameter effects are found to exert an important influence at that stage. There are indeed several parameters to take into account such as the scaled mass, moment of inertia, gravity, along with the scaled initial conditions. The final section \ref{sec:conc} provides a further discussion.

\section{The fluid-body interaction}\label{sec:fbi}

\begin{figure}
\centering
\subfloat[\label{fig:config}]{\includegraphics[scale=0.4]{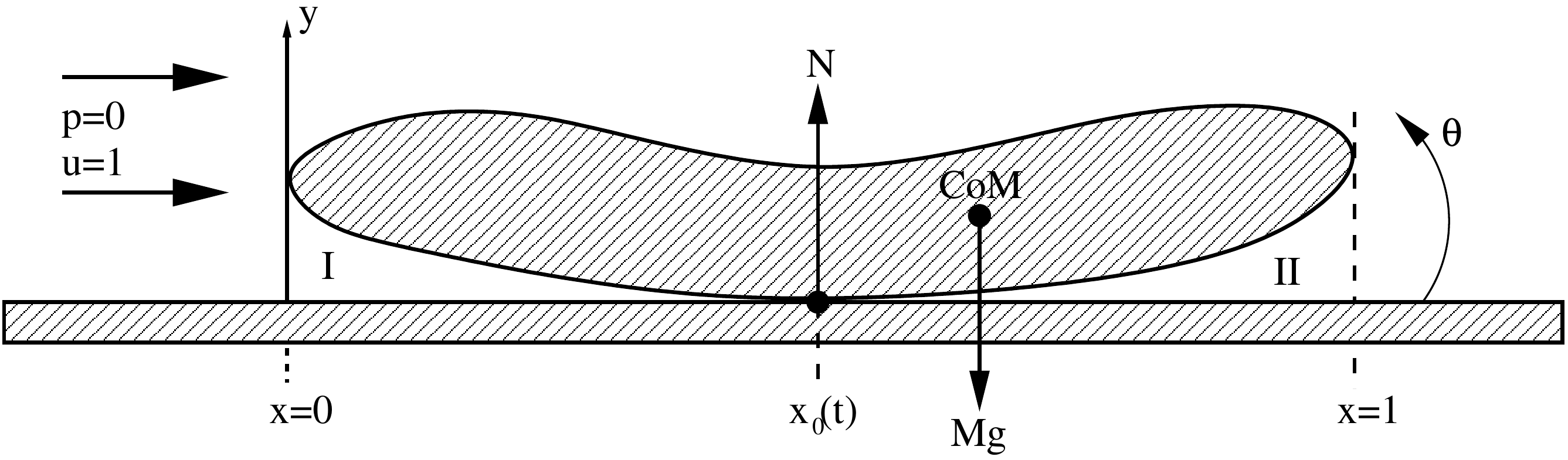}}\\
\subfloat[\label{fig:shapes}]{\includegraphics[scale=0.2]{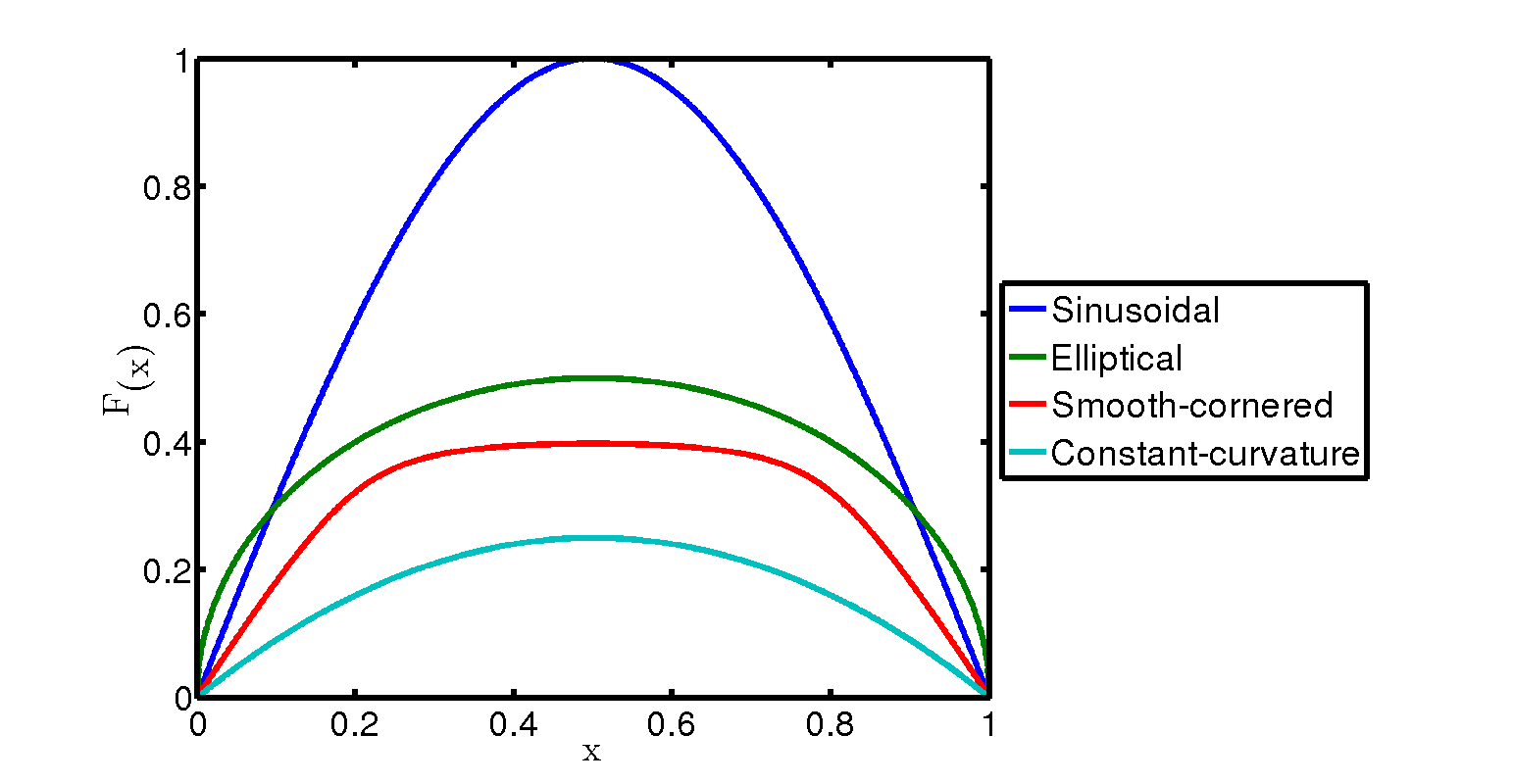}}
\caption{(\textit{a}) A diagrammatic sketch of the flow structure showing the main two thin layers I, II under the rocking body, the fixed centre of mass (CoM) at $x=x_c$, the contact point $x=x_0(t)$ at the current time $t$, and the oncoming stream of fluid. Here in Cartesian coordinates, $x$ is horizontal and $y$ is vertically upwards. (\textit{b}) The four representative body under-surface shapes considered herein. The sinusoidal body has $F(x)=\sin(\pi x)$; the elliptical has  $F(x)=\sqrt{x-x^2}$; the smooth-cornered has $F(x)=-\sqrt{X_a(x)} - \sqrt{X_b(x)} + \sqrt{X_a(0)} + \sqrt{X_b(x)}$, where $X_a(x)=(x-a_1)^2+a_2^2$, $X_b(x)=(x-b_1)^2+b_2^2$, and $(a_1,a_2,b_1,b_2)=(0.2,0.1,0.8,0.1)$; and the constant-curvature has $F(x)=x(1-x)$.}
\end{figure}

The interaction between the body and the surrounding fluid is described as follows, with non-dimensional variables being used throughout.  Cartesian velocity components $(u,v)$ and pressure $p$ are based on the horizontal incident fluid speed $U_D$ and $\rho_D U_D^2$, respectively, while the corresponding Cartesian coordinates $(x,y)$ and time $t$ are based on the characteristic body length $L_D$ and travel time $L_D/U_D$ in turn. The oncoming flow is thus a uniform stream with $u=1$, $v=0$. The body is thin, being of typical scale $O(\bar{h})$, $\bar{h}\ll 1$, in $y$ and length unity in $x$, and is taken to be of smooth shape. In the configuration diagram of figure \ref{fig:config}, the stream is depicted as moving fluid from left to right. The surface, or wall, upon which the body rests is horizontal. The body is in contact with the wall, at least in the first instance, at a single contact point as shown. An inviscid or quasi-inviscid theory is applied here, with separation-free unsteady flow assumed in the two thin fluid-filled gaps labelled I,II on either side of the contact point, whose position varies with time.

In the left-hand gap I ahead of the unknown contact point at $x=x_0(t)$, we have in non-dimensional terms
\begin{subequations}\label{eq:gap1full}
\begin{align}
\label{eq:gap1h} H_t+(uH)_x &=0\ ,\\
\label{eq:gap1u} u_t+uu_x &=-p_x\ ,\\
\label{eq:gap1p} p+\frac{1}{2}u^2 &= \frac{1}{2} \quad\text{at } x=0\ ,\\
\label{eq:gap1attach} u &=x_0'(t) \quad\text{at } x=x_0(t)\ .
\end{align}
\end{subequations}
Subscripts $t,x$ denote partial derivatives, whereas the prime denotes an ordinary derivative with respect to the relevant variable, in this case time $t$. The condition (\ref{eq:gap1p}) allows for a jump across the leading-edge Euler zone, with the leading edge itself being at $x=0$ and with inflow being supposed locally. The attachment property (\ref{eq:gap1attach}), which is discussed in some detail later, is associated with smoothness of the local flow solution close to the contact point.

In the right-hand gap II downstream of the contact point we have similarly the governing equations
\begin{subequations}\label{eq:gap2full}
\begin{align}
\label{eq:gap2h} H_t+(uH)_x &=0\ ,\\
\label{eq:gap2u} u_t+uu_x &=-p_x\ ,\\
\label{eq:gap2p} p &= 0 \quad\text{at } x=1\ ,\\
\label{eq:gap2attach} u &=x_0'(t) \quad\text{at } x=x_0(t)\ ,
\end{align}
\end{subequations}
differing from (\ref{eq:gap1full}) only in (\ref{eq:gap2p}). This condition at the effective trailing edge $x=1$ in II is appropriate to the external-flow configuration, with atmospheric pressure being taken as zero for convenience. The local flow in the thin gap is supposed to be outward. The condition also applies, however, to internal flows when the gap on one side of the body (in this case, the lower side) is small relative to that on the other side, as in \citet{smithellis10}. The reason for the requirement (\ref{eq:gap2p}) is that under the present assumption of a long thin body the pressure varies typically by only a small amount of order $\bar{h}$ throughout the external flow compared with its characteristic $O(1)$ variation within the two gaps as shown in (\ref{eq:gap1full}a--d, \ref{eq:gap2full}a--d). The same requirement results if the part of the body exposed to external flow is bluff and preserves the separation-free motion.

Coupled with the fluid-flow equations above are the body-motion equations, namely
\begin{eqnarray}
\label{eq:bodyh} Mh''(t) &=& \int_0^1 p(x,t) \ \mathrm{d}x + N(t) - Mg^+\ ,\\
\label{eq:bodyth} I\theta''(t) &=& \int_0^1 (x-x_c)p(x,t)\ \mathrm{d}x + (x_0-x_c)N(t)\ .
\end{eqnarray}
Here $x=x_c$ is the prescribed $x$-location of the centre of mass and $h(t)$ its unknown vertical $y$-location, while $\theta(t)$ is the unknown angle the body chord line makes with the horizontal. The factors $M,I$ are scaled effects of mass and moment of inertia, while $g^+$ is the scaled acceleration due to gravity, and $N(t)$ is the unknown scaled normal reaction force acting on the body due to contact with the wall. The dimensional mass and moment of inertia are $\rho_DL_D^2M/\bar{h}$, $\rho_DL_D^4I/\bar{h}$ respectively, the dimensional gravity is $\bar{h}U_D^2g^+/L_D$ (the Froude number is thus $1/(\bar{h}g^+)$ while the Richardson number is $\bar{h}g^+$) and the dimensional normal reaction force is $\rho_DU_D^2L_DN$. In formulating the body-motion equations (\ref{eq:bodyh},\ref{eq:bodyth}) we have supposed that $N(t)$ is positive; if it should ever turn out that $N$ becomes negative then the body will be taken to be no longer in contact with the wall. The marginal case $N=0$ remains moot.  The integral contributions in  (\ref{eq:bodyh},\ref{eq:bodyth}) are dominated by the gap pressures for the assumed thin body but if the external part of the body is bluff instead then the latter part also contributes. An assumption of integrability also needs to be mentioned with regard to integration through the $x_0$ contact point in (\ref{eq:bodyh},\ref{eq:bodyth}). In addition, the unknown gap shapes upstream and downstream of the contact point are given by
\begin{equation}\label{eq:Hdef}
H(x,t) = h(t) + (x-x_c)\theta(t)-F(x)\ ,
\end{equation}
where $F(x)$ is the prescribed smooth shape of the under-surface of the body. The four main shapes considered in this paper are shown in figure \ref{fig:shapes}. Both $h$ and $\theta$ are functions of $t$ which are unknown in advance. Clearly at the contact point the constraints
\begin{equation}\label{eq:smooth}
H=0\ ,\frac{\partial H}{\partial x}=0 \quad \text{at } x=x_0(t)
\end{equation}
hold for the smooth shapes considered herein.

Our task in general is to solve the nonlinear system (\ref{eq:gap1full})--(\ref{eq:smooth}) for $u,p,h,\theta$ in effect. Of interest first is the behaviour just after the motion starts at time $t=0$, say. An analysis of the fluid-body interaction then, as presented in the following section, is found to complement the subsequent numerical work as well as to yield helpful results concerning the physical understanding of the nonlinear interaction present.

\section{Small-time properties}\label{sec:smallt}

The body is supposed to be positioned initially with its contact point with the wall being at some station $x_0=A$ say when the stream is suddenly switched on at time zero.  This is equivalent to abrupt application of a fluid flow or a significant change in the fluid flow.  The body motion itself is assumed to start from rest. A first guess for the response at small positive times is that the constant-pressure form where $p=1/2$ throughout layer I, with $p=0$ in II, might work as an initial condition with zero initial flow in each layer, since then (\ref{eq:gap1u},c), (\ref{eq:gap2u},c) are all satisfied. The guess is then modified, however, by (\ref{eq:bodyh},\ref{eq:bodyth}) requiring the variation in $h,\theta$ to be of order $t^2$, implying a variation of order $t^2$ in the gap thickness $H$ from (\ref{eq:Hdef}), which leads on to $u$ being of order $t$ via (\ref{eq:gap1h}), and hence the pressure variation is of order unity. As a result, we expect the small-time expansions to take the form
\begin{subequations}
\begin{align}
p &= p_0(x) + tp_1(x) + t^2p_2(x)+\dots\ ,\\
u &= tu_1(x) + t^2u_2(x)+\dots
\end{align}
\end{subequations}
in layer I, with the coefficients being functions of $x$ to be determined, and similarly
\begin{subequations}
\begin{align}
P &= P_0(x) + tP_1(x) + t^2P_2(x)+\dots\ ,\\
U &= tU_1(x) + t^2U_2(x)+\dots
\end{align}
\end{subequations}
within layer II. The corresponding body movement has 
\begin{subequations}
\begin{align}
h(t) &= h_0 + t^2h_2+\dots\ ,\\
\theta(t) &= \theta_0 + t^2\theta_2+\dots\ ,\\
x_0(t) &= A + tB + t^2C + \dots\ ,
\end{align}
\end{subequations}
where $0<A<1,B,C$ are constants, but inspection of the requirement (\ref{eq:gap1attach}) for attachment soon establishes that $B=0$. Here the initial state satisfies
\begin{subequations}\label{eq:bodyinit}
\begin{align}
h_0-F(A)+(A-x_c)\theta_0 &=0\ ,\\
F'(A) &=\theta_0\ ,
\end{align}
\end{subequations}
by virtue of the contact condition (\ref{eq:smooth}).

Substitution into (\ref{eq:smooth}) now shows at order $t^2$ that the shape effects $h_2,\theta_2,C$ must be connected by the two relations
\begin{subequations}
\begin{align}
\label{eq:uprot} h_2+(A-x_c)\theta_2 &=0 \ ,\\
\label{eq:c} CF''(A) &=\theta_2\ ,
\end{align}
\end{subequations}
if the smooth body remains on the wall. An interpretation of (\ref{eq:uprot}) can be made in terms of a combined upward and rotational movement of the body at small times.

Meanwhile, the body motion in (\ref{eq:bodyh},\ref{eq:bodyth}) requires at leading order the balances
\begin{subequations}\label{eq:no}
\begin{align}
2Mh_2 &= J_1 + N_0 - Mg^+\ ,\\
\label{eq:noI} 2I\theta_2 &= J_2 + (A-x_c)N_0
\end{align}
\end{subequations}
to hold, where $N_0$ is the leading $O(1)$ contribution to the reaction force $N$, and $J_1,J_2$ are defined as
\begin{subequations}\label{eq:j1j2}
\begin{align}
J_1 &= \int_0^A p_0(x)\ \mathrm{d}x + \int_A^1 P_0(x)\ \mathrm{d}x\ ,\\
\label{eq:j2} J_2 &= \int_0^A (x-x_c)p_0(x)\ \mathrm{d}x + \int_A^1 (x-x_c)P_0(x)\ \mathrm{d}x\ .
\end{align}
\end{subequations}
Thus by elimination of $N_0$ a relation between $h_2,\theta_2$ is inferred, namely
\begin{subequations}\label{eq:h2th2}
\begin{align}
\alpha_2 h_2 + \beta_2\theta_2 &= \gamma_2\ ,\\
\intertext{with}
\alpha_2 &= -2M(A-x_c)\ ,\\
\beta_2 &= 2I\ ,\\
\gamma_2 &= J_2 + (A-x_c)(Mg^+-J_1)\ .
\end{align}
\end{subequations}
The coefficients $\alpha_2,\beta_2$ are known constants,  whereas $\gamma_2$ clearly depends through $J_1,J_2$ on integral properties of the unknown pressure coefficients $p_0(x),P_0(x)$: this represents an influence of added mass. We must move on to determine those pressure coefficients in terms of $h_2$, or $\theta_2$, or both.

In the fluid flow in layer I, from (\ref{eq:gap1h}) at the order of $t$ we obtain an equation controlling $u_1(x)$, given $H_0(x)=h_0+(x-x_c)\theta_0 - F(x)$ as the known initial gap shape. Integration in $x$ then leads to 
\begin{equation}
u_1(x)H_0(x) = -2\left(h_2 + \left(\frac{x+A}{2} - x_c\right)\theta_2\right)(x-A)\ .
\end{equation}
A requirement of finiteness for $u_1$ at the original contact position $x=A$ is imposed in order to keep the associated pressure coefficient finite at contact in line with the fluid-body interaction structure. In fact, (\ref{eq:gap1u}) at leading order indicates that the induced pressure gradient remains finite at $x=A-$ since $h_2+(x-x_c)\theta_2$ is then of $O(x-A)$ from (\ref{eq:uprot}), while the gap shape $H_0(x)$ is then of $O(x-A)^2$ in view of (\ref{eq:bodyinit}). The results now yield
\begin{subequations}
\begin{align}
u_1(x) &= -\frac{\theta_2 (x-A)^2}{H_0(x)}\ ,\\
p_0(x) &= \theta_2\int_A^x \frac{(x-A)^2}{H_0(x)}\ \mathrm{d}x + p_0(A-)\ ,
\end{align}
\end{subequations}
where it is noted in particular that the integral is convergent for all $x$. Furthermore, the leading-edge condition (\ref{eq:gap1p}) becomes $p_0(0)=1/2$ here, which relates $p_0(A)$ to $\theta_2$ and leaves us with
\begin{subequations}\label{eq:p0}
\begin{align}
\label{eq:p0eq}p_0(x) &= \frac{1}{2}+\theta_2K_1\ ,\\
\text{where } K_1(x) &= \int_0^x \frac{(x-A)^2}{H_0(x)}\ \mathrm{d}x\ ,
\end{align}
\end{subequations}
for $0<x<A$. Exactly the same approach applies in layer II except for $P_0$ replacing $p_0$ and the trailing-edge condition (\ref{eq:gap2p}) replacing (\ref{eq:gap1p}). Therefore we find
\begin{subequations}\label{eq:P0}
\begin{align}
\label{eq:P0eq} P_0(x) &= \theta_2K_2\ ,\\
\text{where } K_2(x) &= \int_1^x \frac{(x-A)^2}{H_0(x)}\ \mathrm{d}x\ ,
\end{align}
\end{subequations}
for $A<x<1$. The functions $K_1,K_2$ are specified functions of $x$.

Hence returning to (\ref{eq:j1j2},\ref{eq:h2th2}) we have now
\begin{equation}\label{eq:h2th2hat}
\alpha_2 h_2 + \hat{\beta}_2 \theta_2 = s\ ,
\end{equation}
with $\hat{\beta}_2 = \beta_2-r$ and the constants $r,s$ are
\begin{subequations}
\begin{align}
\nonumber r &= \int_0^A (x-x_c)K_1(x)\ \mathrm{d}x + \int_A^1 (x-x_c)K_2(x)\ \mathrm{d}x\\
\label{eq:r} &\quad - (A-x_c)\left( \int_0^A K_1(x)\ \mathrm{d}x + \int_A^1 K_2(x)\ \mathrm{d}x\right)\ ,\\
s &= \frac{A(A-2x_c)}{4} + (A-x_c)\left(Mg^+ - \frac{A}{2}\right)\ .
\end{align}
\end{subequations}

The main system to be tackled is therefore (\ref{eq:uprot}),(\ref{eq:h2th2hat}) for $h_2,\theta_2$. Afterwards $C$ is determined by (\ref{eq:c}), and we can in addition use (\ref{eq:no}) to check on whether the primary reaction-force contribution $N_0$ is positive or negative. It is noteworthy here that the contact-attachment requirement (\ref{eq:gap1attach}) at the present level of approximation reprodcues the smooth shape condition (\ref{eq:c}), since the curvature effects $H_0''$ and $-F''$ are identical. Moreover, the solutions for $p_0,P_0$ in (\ref{eq:p0},\ref{eq:P0}) indicate that there is an $O(1)$ pressure jump produced across the contact position.

\begin{figure}
\centering
\subfloat[\label{fig:smallta}]{\includegraphics[scale=0.16]{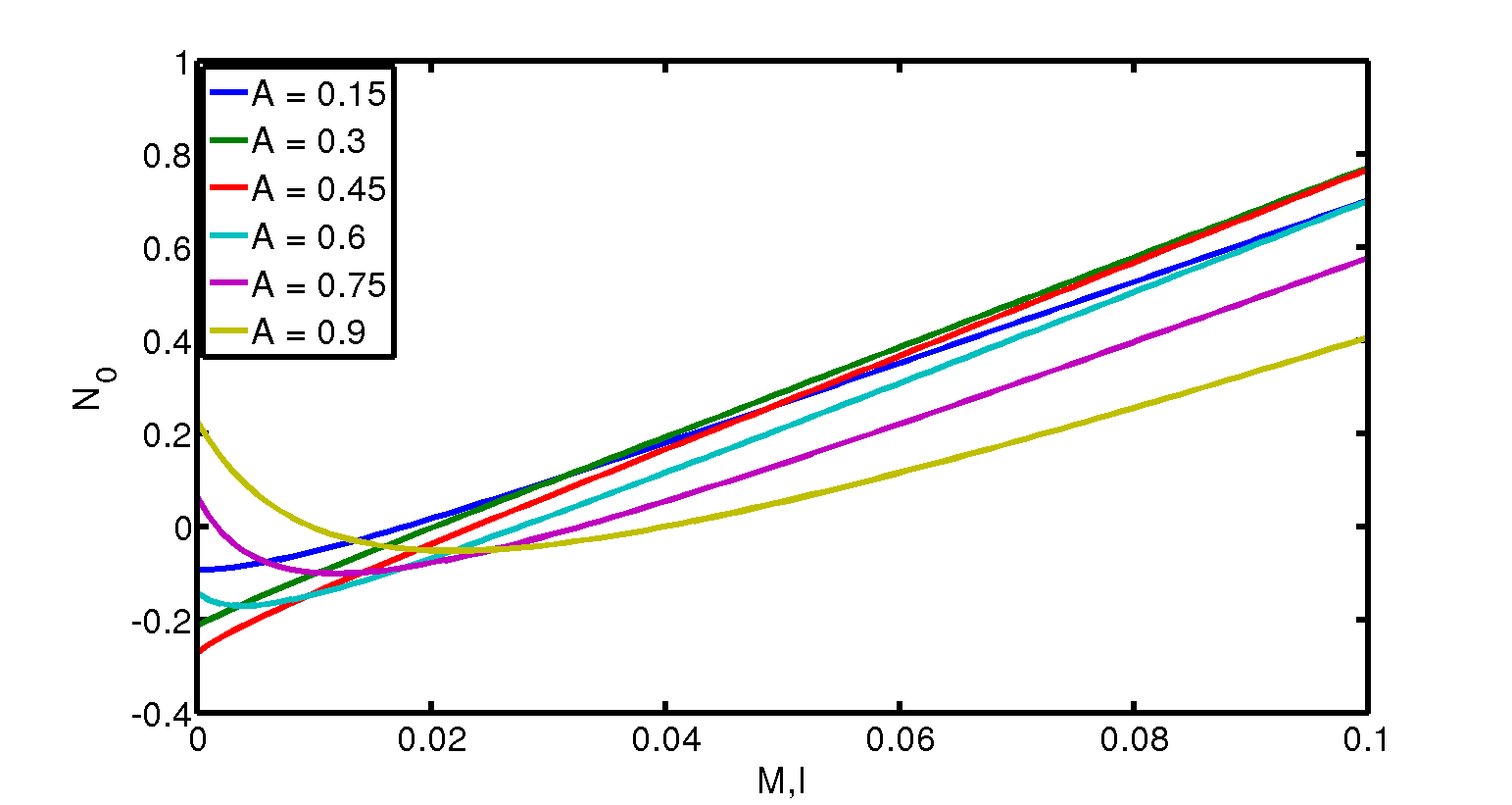}}
\subfloat[\label{fig:smalltb}]{\includegraphics[scale=0.16]{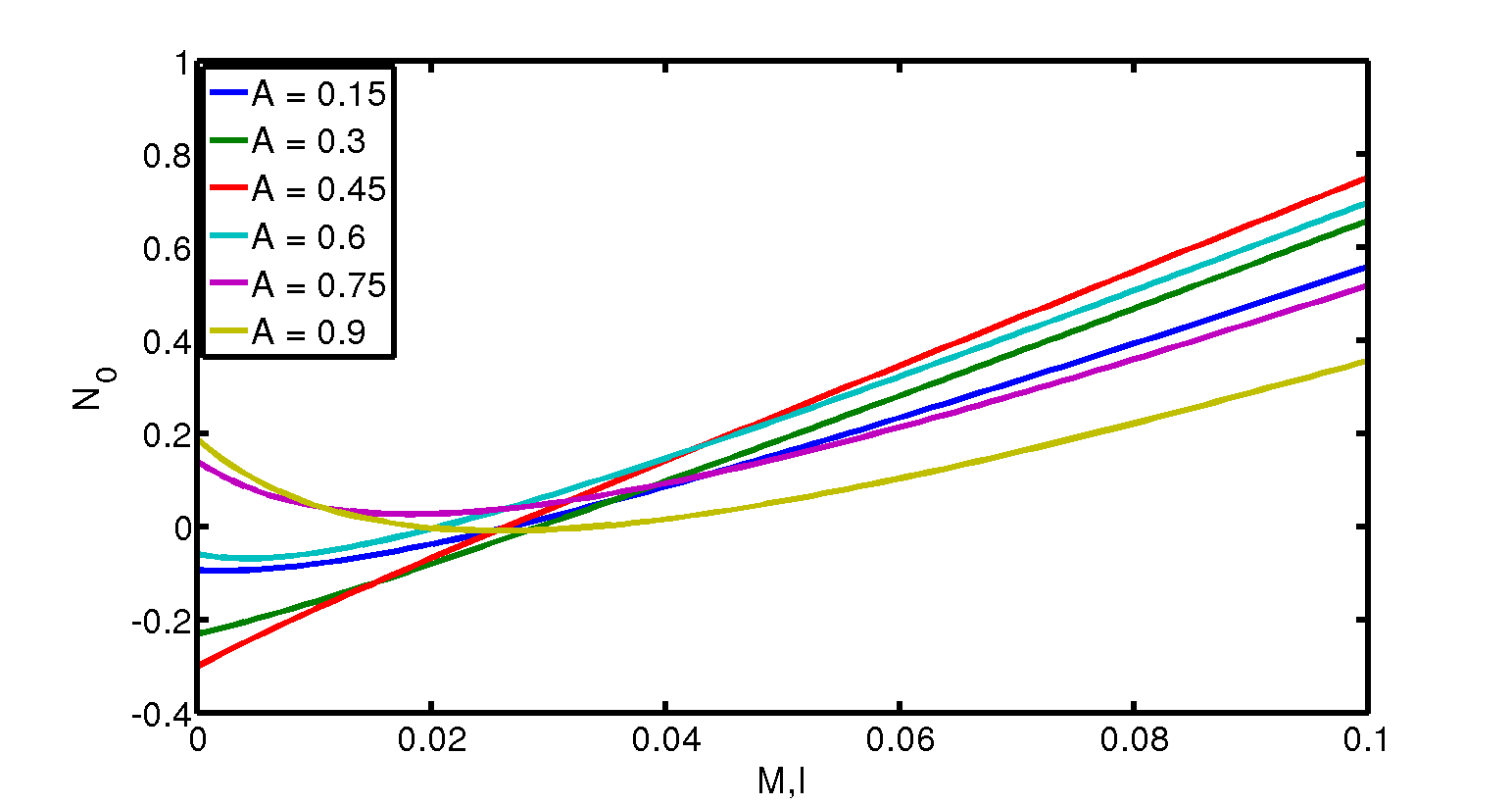}}\\
\subfloat[\label{fig:smalltc}]{\includegraphics[scale=0.16]{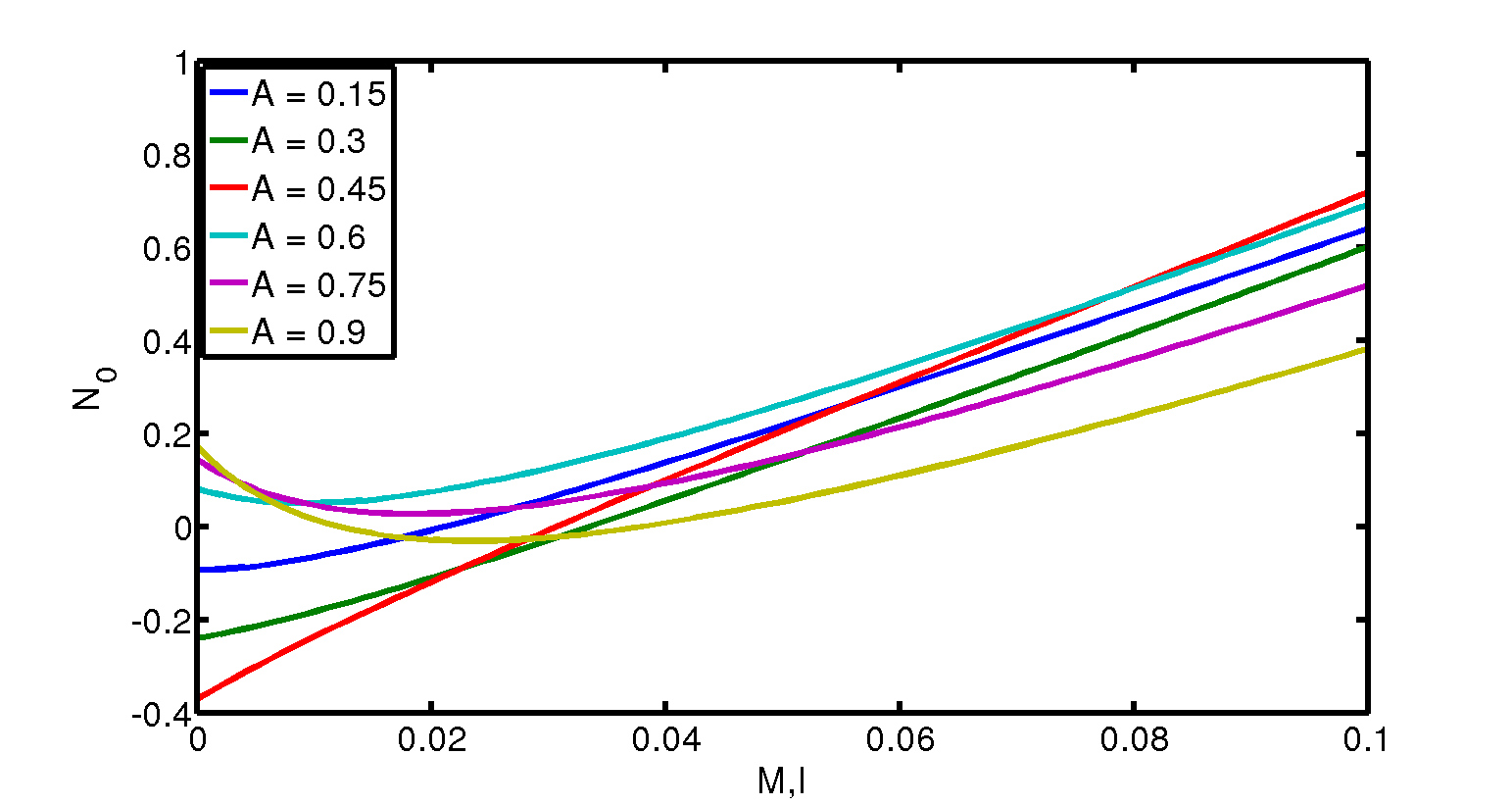}}
\subfloat[\label{fig:smalltd}]{\includegraphics[scale=0.16]{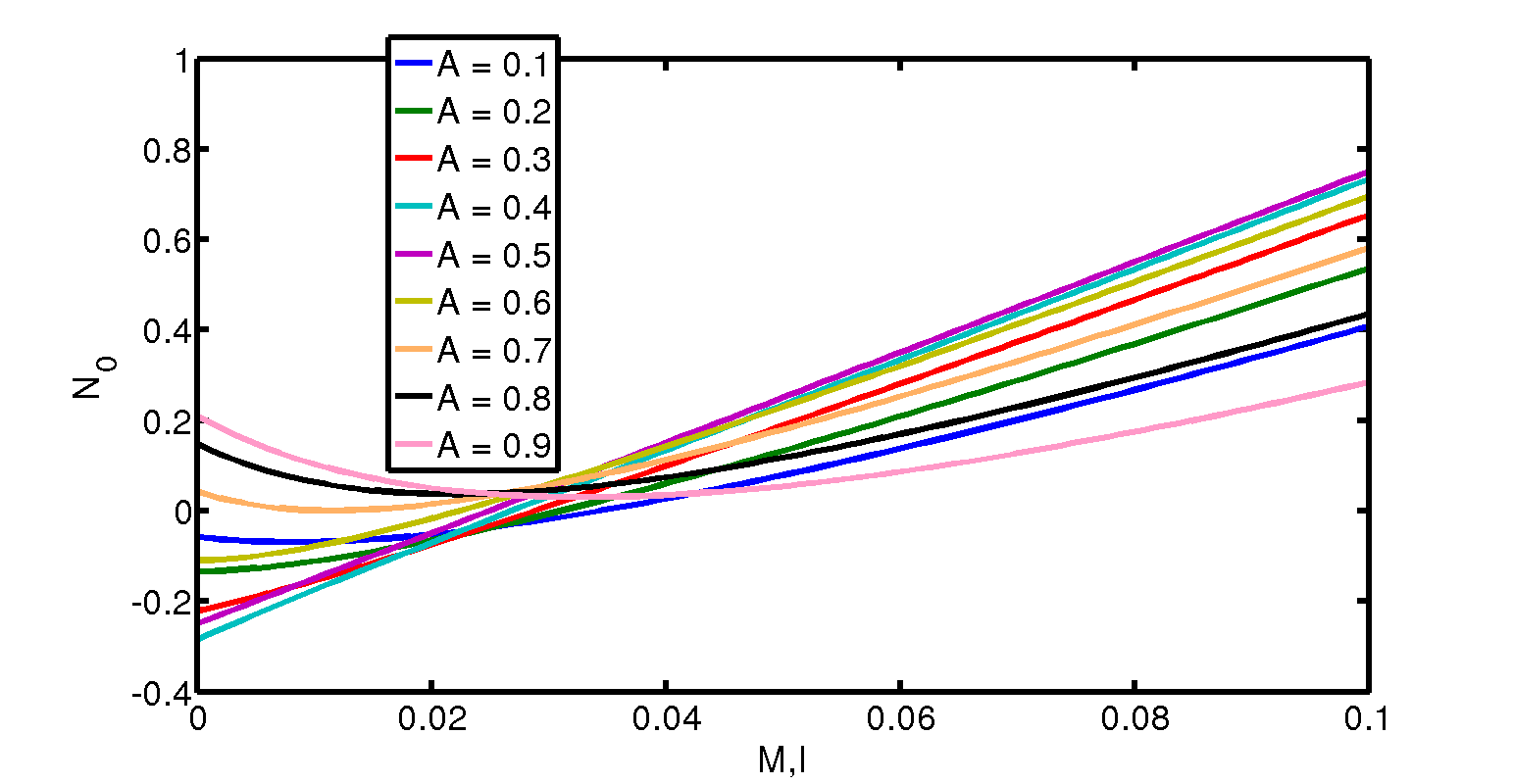}}\\
\subfloat[\label{fig:smallte}]{\includegraphics[scale=0.16]{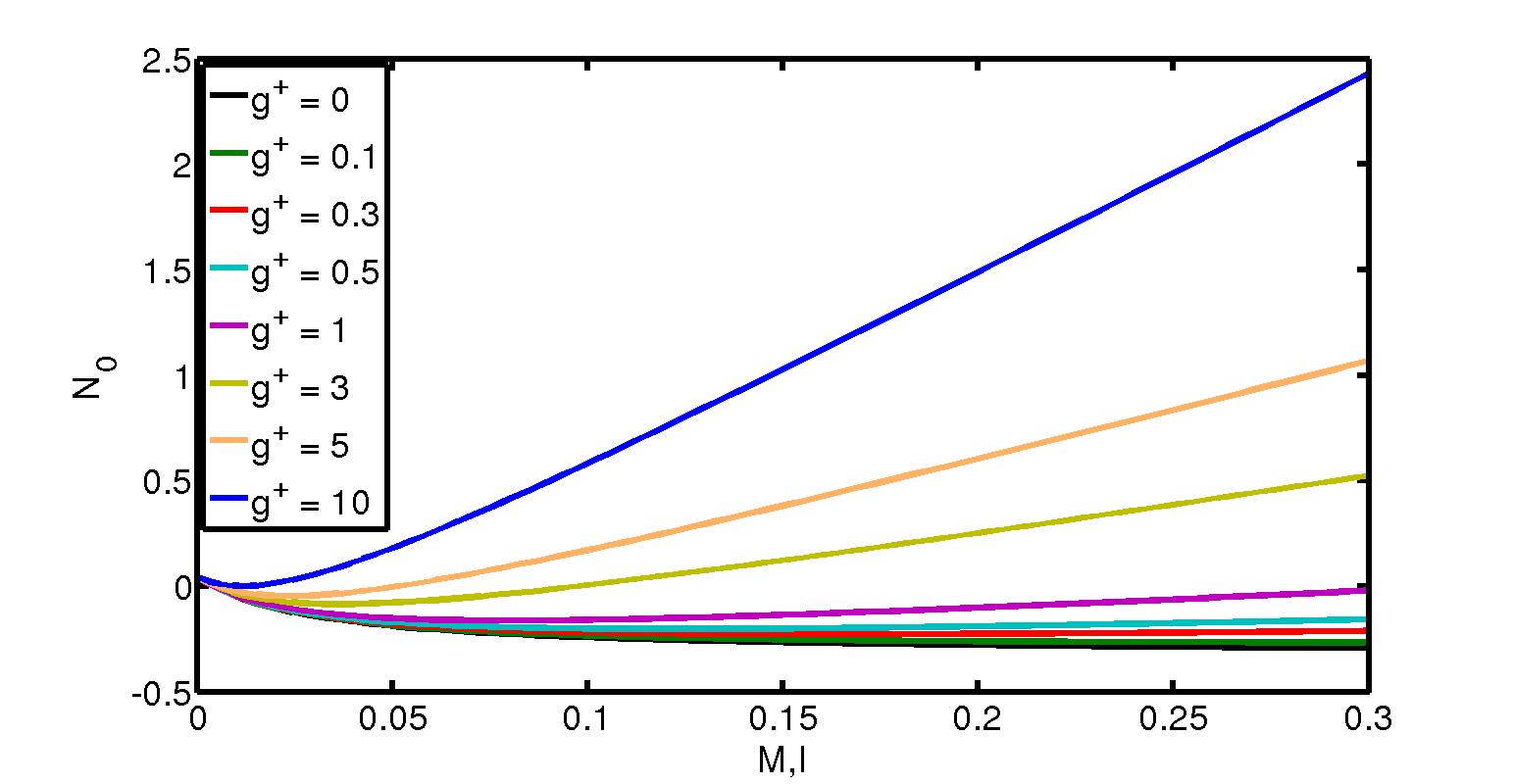}}
\subfloat[\label{fig:smalltf}]{\includegraphics[scale=0.16]{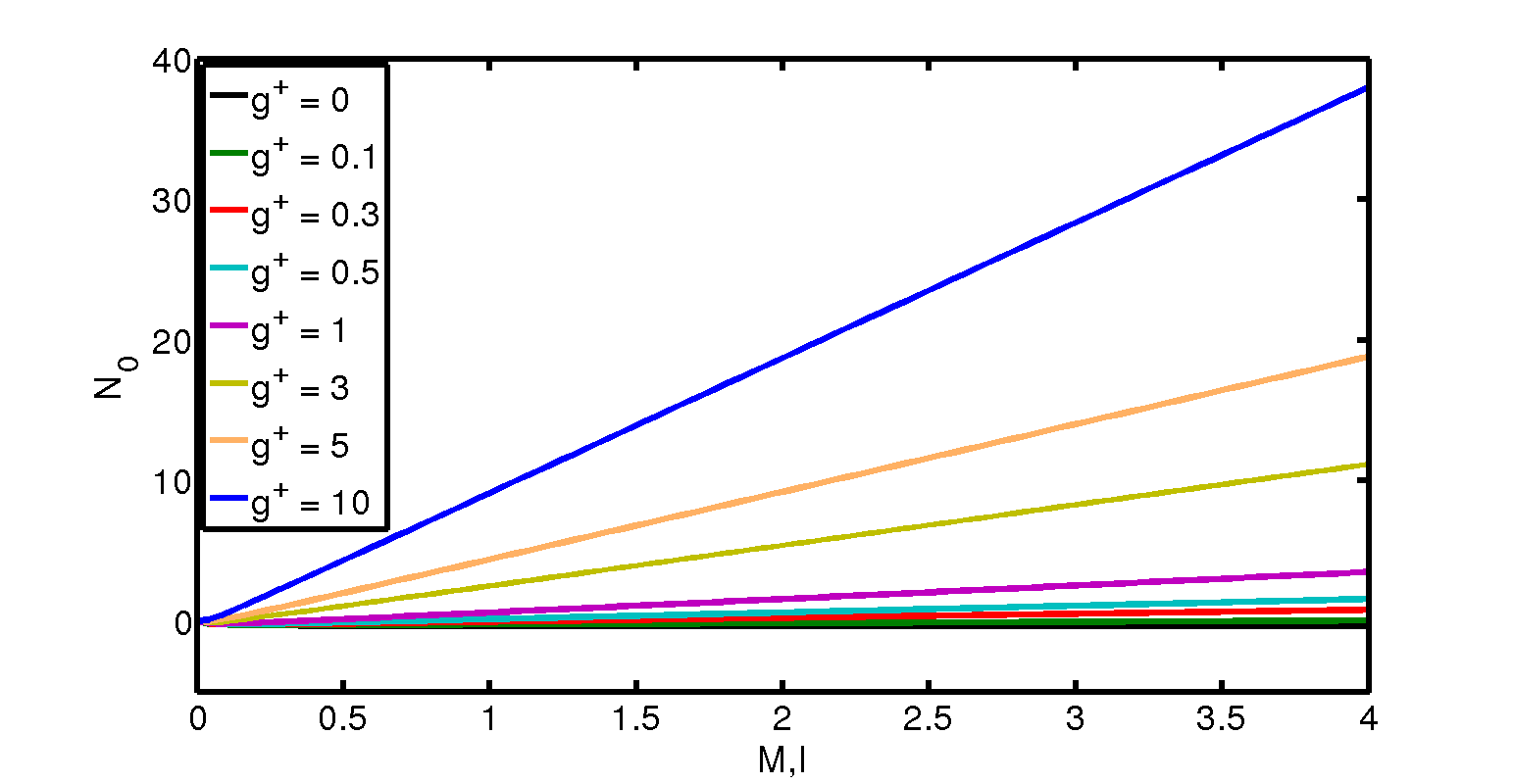}}
\caption{Numerical solutions for the small-time system (\ref{eq:uprot},b),(\ref{eq:h2th2hat}). Each figure shows for different body shapes $F(x)$ the scaled leading-order vertical reaction force $N_0$ versus the scaled body mass $M$ for various values of the parameter $A$ governing the initial location. Here, $I=M$ for definiteness. (\textit{a}) Sinusoidal body $F(x)=\sin(\pi x)$, $x \in (0,1)$; $g^+=10$. (\textit{b}) Elliptical body $F(x)=\sqrt{x-x^2}$, $x \in (0,1)$; $g^+=10$. (\textit{c}) Smooth-cornered body $F(x)=-\sqrt{X_a(x)} - \sqrt{X_b(x)} + \sqrt{X_a(0)} + \sqrt{X_b(x)}$, where $X_a(x)=(x-a_1)^2+a_2^2$, $X_b(x)=(x-b_1)^2+b_2^2$, and  $(a_1,a_2,b_1,b_2)=(0.2,0.1,0.8,0.1)$, $x \in (0,1)$; $g^+=10$. (\textit{d}) Constant curvature body $F(x)=x(1-x)$, $x \in (0,1)$; $g^+=10$. (\textit{e}) The influence of the (inverse) Froude number on lift-off. Here, the body shape is the constant curvature case of figure \ref{fig:smalltd} with $A= 0.7$, but $g^+$ is varied from 10 down to zero. (\textit{f}) As for figure \ref{fig:smallte} but over a wider range of values of M, I. 
}\label{fig:smallt}
\end{figure}

We show the prime features of the small-time solutions in figure \ref{fig:smallt} for the representative body shapes shown in figure \ref{fig:shapes}. Some conditions are found to yield the reaction-force contribution $N_0$ being negative, which corresponds to the occurrence of an immediate lift-off. For an (under-) body shape which is sinusoidal, as in $F(x)=\sin(\pi x)$ between $x$ zero and $x$ unity, figure \ref{fig:smallta} shows the results for the scaled leading-order vertical reaction force $N_0$ versus the scaled body mass $M$ as the parameter $A$ governing the initial location is varied. Here the scaled moment of inertia $I$ is taken to be equal to the scaled mass $M$ for definiteness; keeping the ratio $I/M$ fixed is meaningful in the sense that the body shape remains fixed but  the dimensional density and relative gap width can still be varied for example. For the horizontally symmetrical shapes investigated in this paper a value $A$ greater than $1/2$ means the body's leading edge is raised above the trailing-edge height.  The results indicate that for all the values $A$ studied there is a finite range of values $M(=I)$ for which $N_0$ becomes negative, implying that for such values the body immediately lifts off from the wall. On the other hand, as $M$ becomes relatively large, the value of $N_0$ always becomes positive and increases linearly with $M$, in keeping with asymptotic behaviour concerning the loss of fluid-flow effects then; the interval within which fluid-flow effects do matter is actually quite small, in this instance being confined to $M$ less than 0.1 roughly. There is in any case a critical value of scaled mass $M$, corresponding dimensionally to the existence of a critical fluid speed for lift-off for a given dimensional mass. Figure \ref{fig:smalltb} is then for an elliptical shape such that $F(x)=\sqrt{x-x^2}$ for $x$ in $(0,1)$, leaving the body slope and curvature singular at the edge points, whereas the sinusoidal body has finite slope and zero curvature at the leading and trailing edges $x=0,1$. Here almost all values of $A$ studied produce a range of values of $M$ in which lift-off can be inferred because $N_0<0$. The case $A=0.8$ is the only exception among those presented. Similarly, figure \ref{fig:smalltc} which is for a smooth-cornered shape, shows ranges of negative $N_0$ and hence lift-off in almost all cases. Here $F(x)=-\sqrt{X_a(x)} - \sqrt{X_b(x)} + \sqrt{X_a(0)} + \sqrt{X_b(x)}$, where the spatially-varying contributions are given by $X_a(x)=(x-a_1)^2+a_2^2$, $X_b(x)=(x-b_1)^2+b_2^2$, and the constants are $(a_1,a_2,b_1,b_2)=(0.2,0.1,0.8,0.1)$. With comparatively large $M(=I)$ in figures \ref{fig:smalltb},(c) the asymptotes are linear as in figure \ref{fig:smallta}. By contrast, with comparatively small or in effect zero $M$ associated with dominant fluid influences, an explicit form of (\ref{eq:uprot}),(\ref{eq:h2th2hat}) can be found: this predicts that for a shape of constant curvature, for example, the critical of $A$ is 2/3 at zero $M$, a critical value which is broadly in line with the results in figures \ref{fig:smallta}--(c). The critical value is also confirmed by the results of figure \ref{fig:smalltd} which are for the shape $F(x)=x(1-x)$ of constant curvature.

Figures \ref{fig:smalltd}, \ref{fig:smallte} indicate respectively the rather sensitive dependence of lift-off or its absence on the body shape and on the gravity factor. Concerning the latter, although most results given in this study take the value of $g^+$ as 10, we may begin to explore now the influences of the (inverse) Froude number. Figure \ref{fig:smallte} has the value of $A$ remaining at 0.7 but $g^+$ is varied from 10 down to zero. The range of conditions that lead to immediate lift-off is increased considerably with such decreasing of $g^+$, a matter which is pursued further in detail in Appendix \ref{app:froude} and leads into the investigations in later sections on finite-time responses.

\section{The behaviour at $O(1)$ times with negligible fluid effects}\label{sec:ord1tnoflow}

Without significant effects from the fluid flow, the controlling balances in (\ref{eq:gap1full})--(\ref{eq:smooth}) over the time scale of order unity reduce to
\begin{eqnarray}
\label{eq:bodyhtord1nf} Mh''(t) &=& N(t) - Mg^+\ ,\\
\label{eq:bodythtord1nf} I\theta''(t) &=&  (x_0-x_c)N(t)\ ,\\
\label{eq:bigh} H(x,t) &=& h(t) + (x-x_c)\theta(t)-F(x)\ ,\\
\label{eq:smoothagain} H=0 &,&\frac{\partial H}{\partial x}=0 \quad \text{at } x=x_0(t)\ .
\end{eqnarray}
Here the unknowns are $N,x_0,h,\theta$ as functions of time $t$ only. The coupled system above corresponds to the relative mass and moment of inertia being large. An assumption of rolling is also noted. Elimination of $N$ in (\ref{eq:bodyhtord1nf}),(\ref{eq:bodythtord1nf}) leads to the equation
\begin{equation}\label{eq:imtord1nf}
I\theta''(t) = M(x_0-x_c)(h''+g^+)\ ,
\end{equation}
which couples with (\ref{eq:bigh}),(\ref{eq:smoothagain}) for $(x_0,h,\theta)(t)$.

Applying (\ref{eq:smoothagain}) on the other hand yields the following successive relationships between $h,\theta$, and $x_0$ involving the body shape function $F(x)$,
\begin{subequations}
\begin{align}
h(t) &=F(x_0) - (x_0-x_c)F'(x_0)\ ,\\
h'(t) &= -(x_0-x_c)F''(x_0)x_0'\ ,\\
\label{eq:hdd} h''(t) &= -F''(x_0)x_0'^2 - (x_0-x_c)(F'''(x_0)x_0'^2+F''(x_0)x_0'')\ ,
\end{align}
\end{subequations}
\begin{subequations}\label{eq:ththdthdd}
\begin{align}
\theta &= F'(x_0)\ ,\\
\theta' &= F''(x_0)x_0'\ ,\\
\theta'' &= F'''(x_0)x_0'^2  + F''(x_0)x_0''\ .
\end{align}
\end{subequations}
Hence from substitution into (\ref{eq:imtord1nf}) we obtain a nonlinear governing equation for $x_0(t)$ alone, which is
\begin{subequations}\label{eq:tidyup}
\begin{align}
\label{eq:tidyupa} \alpha x_0'' + \beta x_0'^2 = zg^+\ &,\\
\intertext{where}
\alpha = \left(\frac{I}{M} +z^2\right)f''\ ,\quad \beta = \frac{I}{M}f''' + z(f''+zf''')\ &,\quad z=(x_0-x_c)\ ,\quad f=F(x_0)\ .\\
\intertext{The fluid-free cases in (\ref{eq:tidyup}a,b) can be integrated once to the energy form, involving a constant of integration $c_2$ which is fixed by the initial conditions,}
\label{eq:dzdt}
\left(\frac{\mathrm{d}z}{\mathrm{d}t}\right)^2 = 2g^+\frac{zf'-f+c_2}{\left(\frac{I}{M} +z^2\right)f''^2}\ &.
\end{align}
\end{subequations}
for any effective body shape $f(z)$. Further integration to obtain $z(t)$ and hence $x_0(t)$ explicitly is shape-dependent, although there are interesting limiting situations such as for the nearly-cornered shapes addressed elsewhere. It can be shown also that the reaction force $N$ remains one-signed in all these fluid-free cases.

\begin{figure}
\centering
\subfloat[\label{fig:tord1noflowa}]{\includegraphics[scale=0.16]{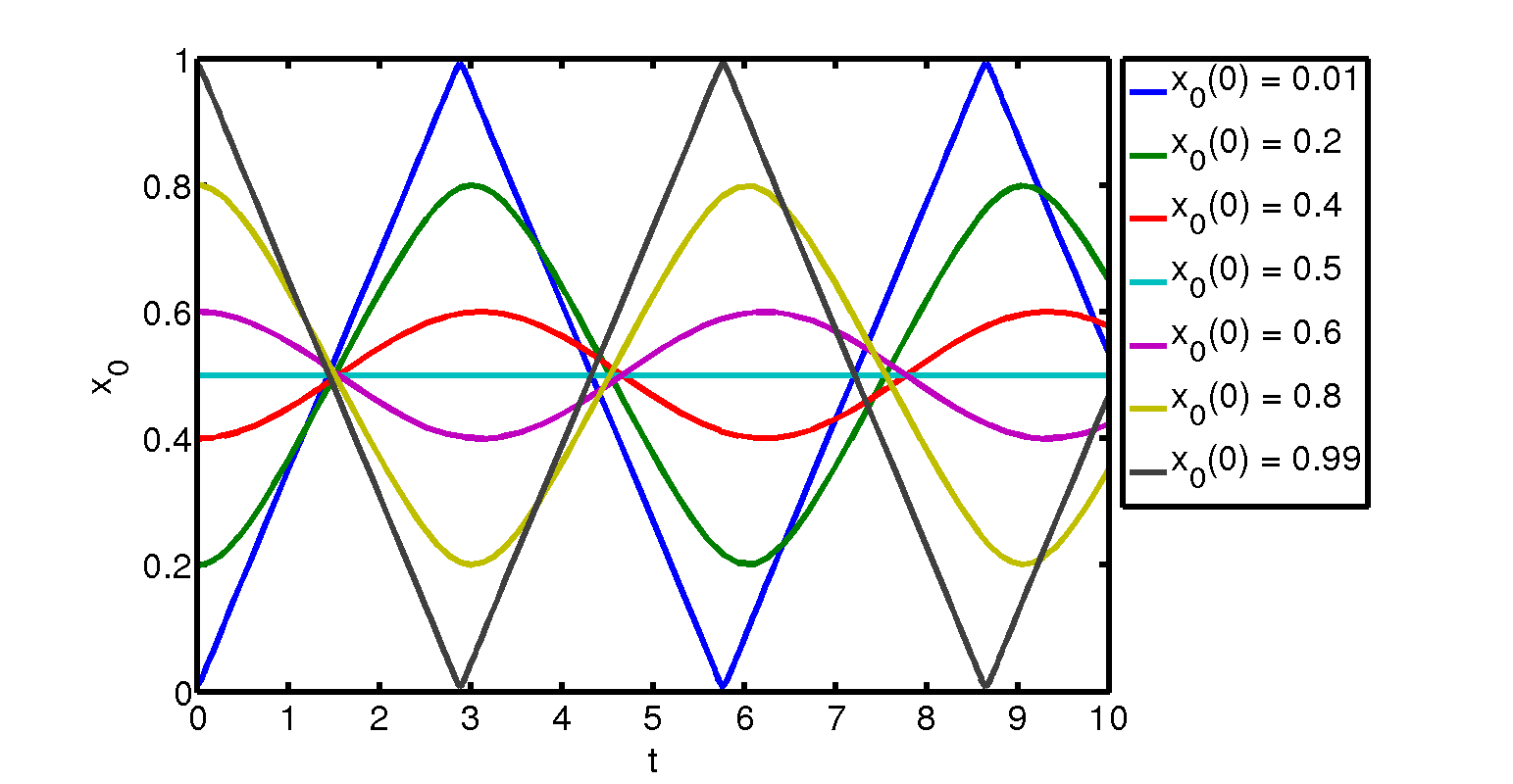}}
\subfloat[\label{fig:tord1noflowb}]{\includegraphics[scale=0.16]{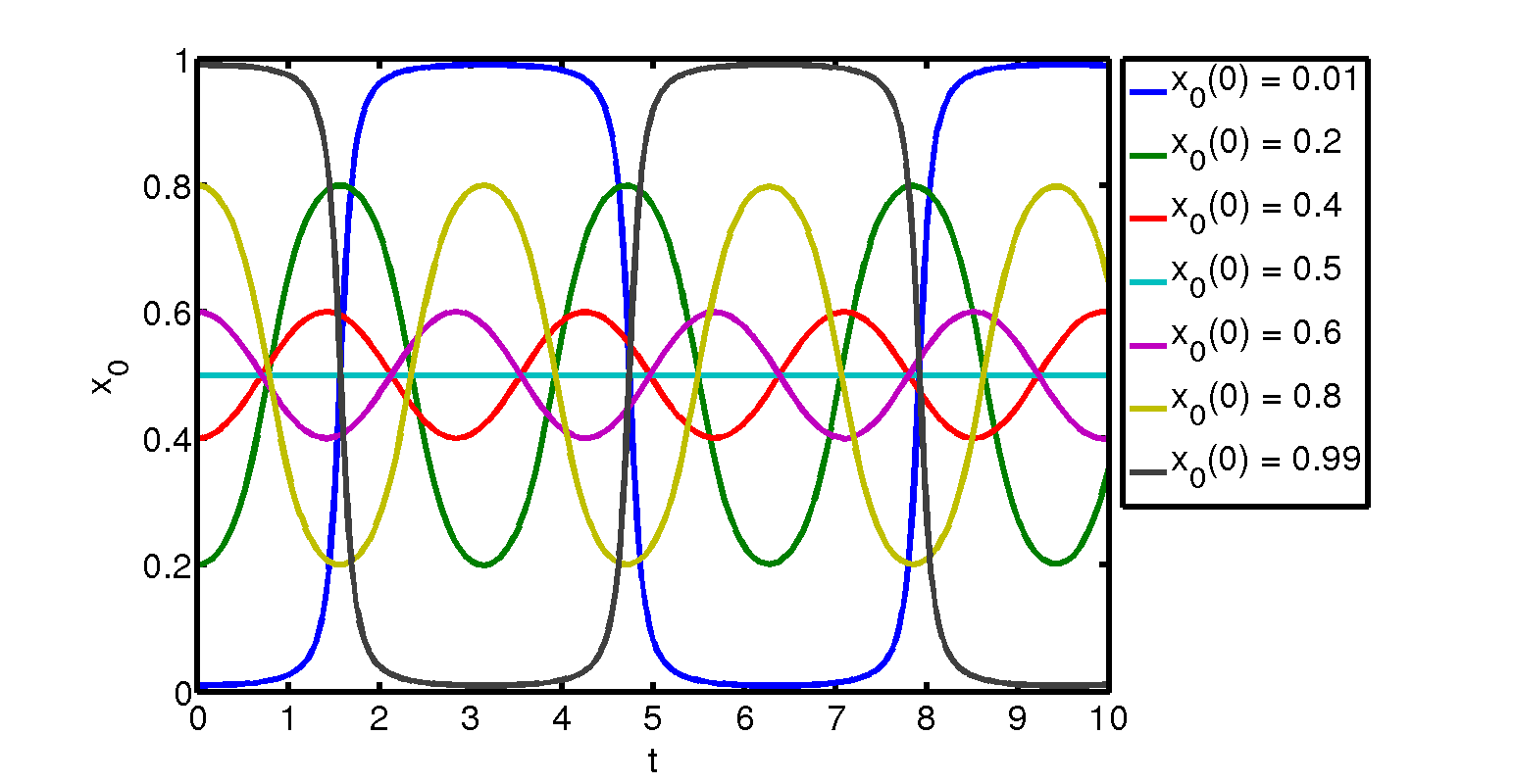}}\\
\caption{Numerical solutions for the case of order unity time and negligible fluid effects (\ref{eq:tidyupa},b) subject to varying initial location of the contact point, while initial contact-point velocity is fixed at zero. In all cases, $g^+=10$, $x_c=0.5$. Figures (\textit{a}) and (\textit{b}) show the time evolution of the contact position $x_0$ for varying initial conditions and body shape: (\textit{a}) is for a sinusoidal body $F(x)=\sin(\pi x)$, $x \in (0,1)$, while (\textit{b}) is for an elliptical body $F(x)=\sqrt{x-x^2}$, $x \in (0,1)$.  }\label{fig:tord1noflow1}
\end{figure}

\begin{figure}
\centering
\subfloat[\label{fig:tord1noflowc}]{\includegraphics[scale=0.2]{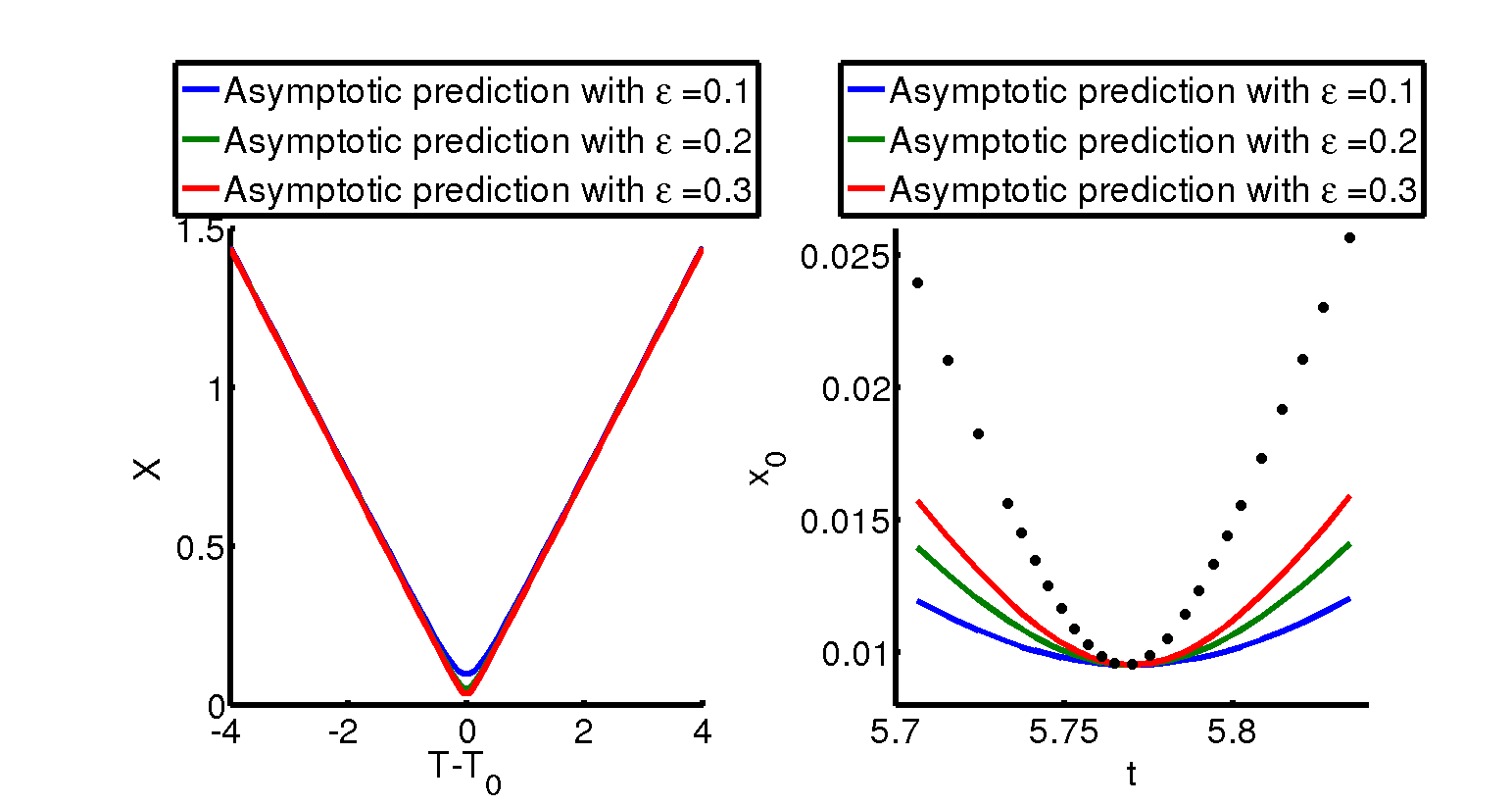}}\\
\subfloat[\label{fig:tord1noflowd}]{\includegraphics[scale=0.2]{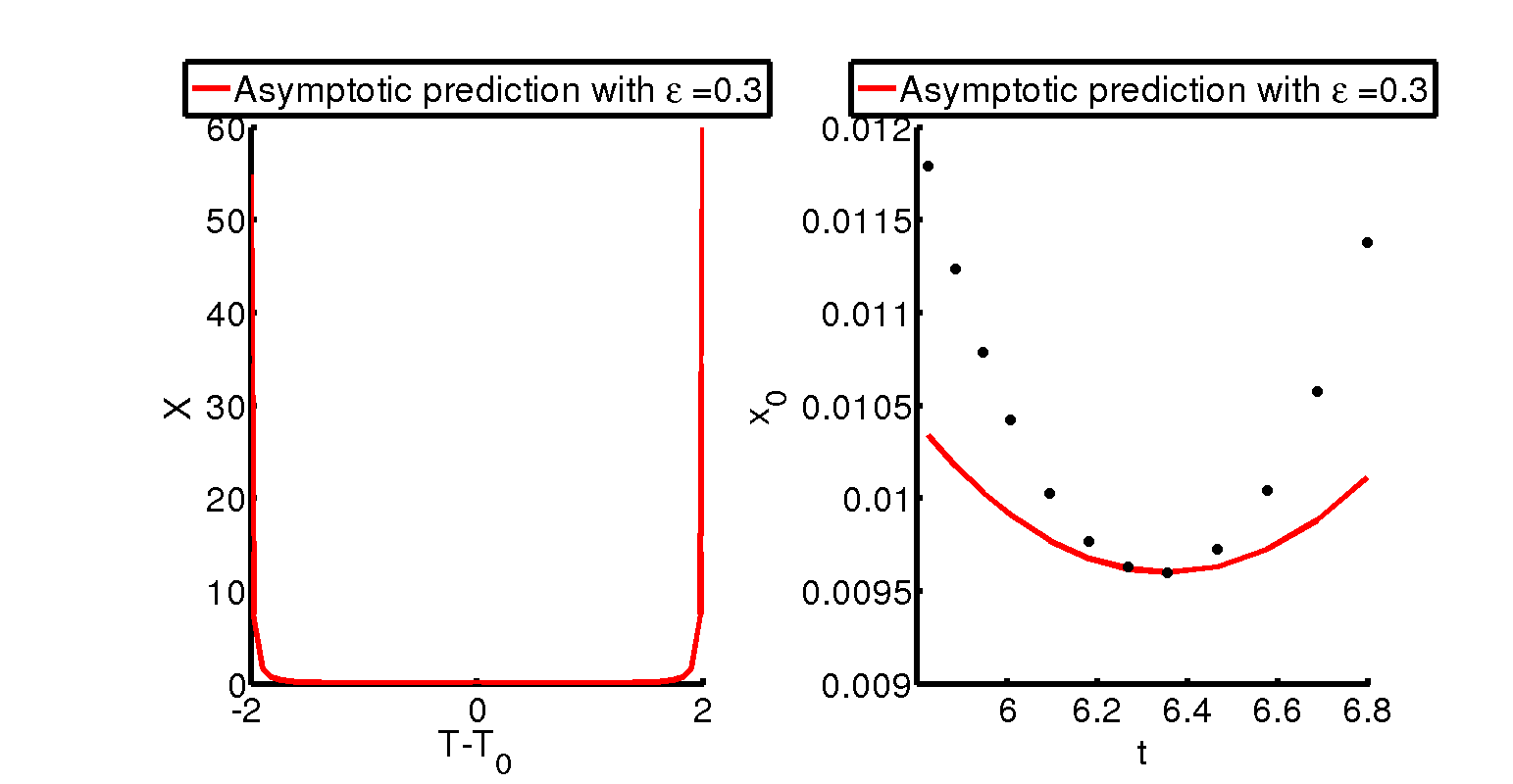}}
\caption{Comparing asymptotic predictions with numerical results in the case of negligible flow effects, and with conditions as in figure \ref{fig:tord1noflow1}. (\textit{a}) The analytical limiting response (\ref{eq:limitsine}) for the sinusoidal body in scaled (left) and unscaled (right) coordinates. The dots in the right-hand figure represent the numerical solutions.  The time $T_0$ at which $x_0$ attains its minimum was obtained from the numerical results in figure \ref{fig:tord1noflowa}. (\textit{b}) As for figure \ref{fig:tord1noflowc}, but for the analytical limiting response (\ref{eq:limitellipse}) for the elliptical body in scaled (left) and unscaled (right) coordinates. Here, a single values of $\epsilon$ is plotted to make the left-hand figure clearer, and since in the right-hand figure the asymptotic predictions from the three values of $\epsilon$ used in figure \ref{fig:tord1noflowc} would  coincide. Once again, dots represent numerical solutions, and the time $T_0$ at which $x_0$ attains its minimum was obtained from the numerical results, as shown in figure \ref{fig:tord1noflowb}.}\label{fig:tord1noflow2}
\end{figure}

Numerical  solutions of (\ref{eq:tidyupa},b), obtained using \textsc{Matlab}'s \textsuperscript{\textregistered} \texttt{ode45} solver, checked by an independent solver, and subject to varying initial conditions, are presented in figures \ref{fig:tord1noflow1}, \ref{fig:tord1noflow2} for two of the earlier prescribed shapes of interest: the sinusoidal shape $F(x)=\sin(\pi x)$ and the elliptical shape $F(x)=\sqrt{x-x^2}$. In both cases, the value of $g^+$ was taken to be 10, while the $x$-position of the centre of mass was fixed as $x_c=0.5$. The sinusoidal shape is shown in figure \ref{fig:tord1noflowa} for initial conditions in which the contact-point velocity $\xi_0=x_0'(0)$ is kept at zero, and the initial contact location $x_0(0)$ is varied over the values shown in the legend. In all cases except that in which $x_0(0)=x_c=0.5$, the solution evolves quite soon into an apparently periodic state in which the body rocks.  The cases $x_0(0)=0.4$ and $x_0(0)=0.6$ remain near equilibrium as $t$ increases with sinusoidal-like oscillations being observed about $x_0=0.5$ in a gentle rocking motion of the body. Values of $x_0(0)$ further away from $x_c$, on the other hand, provoke rapid rocking as these cases come closer to failure or lift-off in the sense that they yield comparatively rapid behavioural changes whenever the contact point $x_0$ approaches either of the end points. However, the second shape, namely the ellipse, is found to be associated with comparatively slow responses whenever the contact point $x_0$ approaches either of the end points, as indicated in figure \ref{fig:tord1noflowb}. Gently-rocking sinusoidal-like oscillations about the equilibrium point at $0.5$ again seem implied for initial conditions $x_0(0)=0.4$ and $x_0(0)=0.6$,  whereas  values of $x_0(0)$ further away from $x_c$ lead to extreme rocking with relatively long periods in which $x_0$ remains close to zero or unity, accompanied by rather rapid increases or decreases of $x_0$ in between. 

Analytical solutions are also possible for certain special shapes but it seems more useful here to examine the limiting responses indicated by the results in figures \ref{fig:tord1noflowa},(b) when the contact point is near an end point. In the locally near-flat scenario of figure \ref{fig:tord1noflowa} the limiting response occurs comparatively rapidly around some time $t=t_0$, such that time $t=t_0+\epsilon T$, say, where the parameter $\epsilon$ is small and $T$ is strictly of order unity, and $x=\epsilon X$ to leading order near the leading edge. In normalised form, the governing equation (\ref{eq:tidyupa}) or (\ref{eq:dzdt}) then leads to the local behaviour of the contact position being given by
\begin{equation}\label{eq:limitsine}
X^2 = \frac{\lambda}{\pi^3}(T-T_0)^2+\Gamma^2 \ ,
\end{equation}
with $\lambda>0,\Gamma>0,T_0$ being the constants $\lambda = g^+x_c((I/M)+x_c^2)^{-1}$, $\Gamma=\epsilon^{-1}x_0$, while $T_0$ is the a priori unknown scaled time at which $x_0$ reaches a minimum.  In the elliptical scenario of figure \ref{fig:tord1noflowb}, by contrast, the time scale is relatively long with $t=\epsilon^{-\frac{1}{4}}T$ and $x=\epsilon X$ again to leading order for small $\epsilon$, so that from (\ref{eq:tidyupa},c) the scaled position $X$ is found to be given by
\begin{equation}\label{eq:limitellipse}
X=\frac{1}{\left(\Gamma^{-1} - \frac{\lambda}{3}(T-T_0)^2\right)^2} \ .
\end{equation}
Here, $\lambda$ and $\Gamma$ are as before, $T_0$ is in general a different value from the sinusoidal case, and $T$ lies between $T_0-\sqrt{3/\lambda\Gamma}$ and $T_0+\sqrt{3/\lambda\Gamma}$ ; close to the singular points there the growth of $X$ matches to the behaviour in the relatively fast transition regions in which $x$ becomes of order unity. Similar accounts apply near the trailing edge, where $1-x=\epsilon X$. The limiting responses just described are presented in figure \ref{fig:tord1noflow2}, and they appear to reflect well the solution properties found in figure \ref{fig:tord1noflow1} corresponding to rapid rocking and long-time rocking, respectively. 

\section{Behaviour at $O(1)$ times with fluid effects}\label{sec:tord1}

Here we address the issue of what happens to the combined nonlinear fluid-body interaction if it persists over the time scale of order unity with significant effects from the fluid flow being present. Instead of (\ref{eq:bodyhtord1nf}), (\ref{eq:bodythtord1nf}) we have now
\begin{eqnarray}
Mh''(t) &=& i_1 + N(t) - Mg^+\ ,\\
I\theta''(t) &=& i_2 +(x_0-x_c)N(t)\ ,
\end{eqnarray}
with $i_1,i_2$ being the integrals from fluid-flow pressure effects in (\ref{eq:bodyh}),(\ref{eq:bodyth}) respectively, given explicitly in (\ref{eq:ize}) below. Elimination of $N$ here leads to the equation
\begin{equation}
I\theta''(t) = i_2 - (x_0-x_c)i_1 + M(x_0-x_c)(h''+g^+)\ ,
\end{equation}
rather than (\ref{eq:imtord1nf}). Further, the successive relationships (\ref{eq:hdd}),(\ref{eq:ththdthdd}) between $h,\theta$, and $x_0$ involving the function $F$ still hold, and so the nonlinear governing equation for $x_0(t)$ now is
\begin{subequations}\label{eq:tord1}
\begin{align}
\label{eq:tord1a} \alpha x_0'' + \beta x_0'^2 &= zg^+ + \frac{i_2-zi_1}{M}\\
\intertext{in this full fluid-body interaction, where, to clarify,}
\label{eq:ize} i_1=\int_0^1 p(x,t)\ \mathrm{d}x \quad &, \quad i_2=\int_0^1 (x-x_c)p(x,t)\ \mathrm{d}x
\end{align}
\end{subequations}
are the flow-pressure contributions.

\begin{figure}
\centering
\subfloat[\label{fig:tord1sinea}]{\includegraphics[scale=0.25]{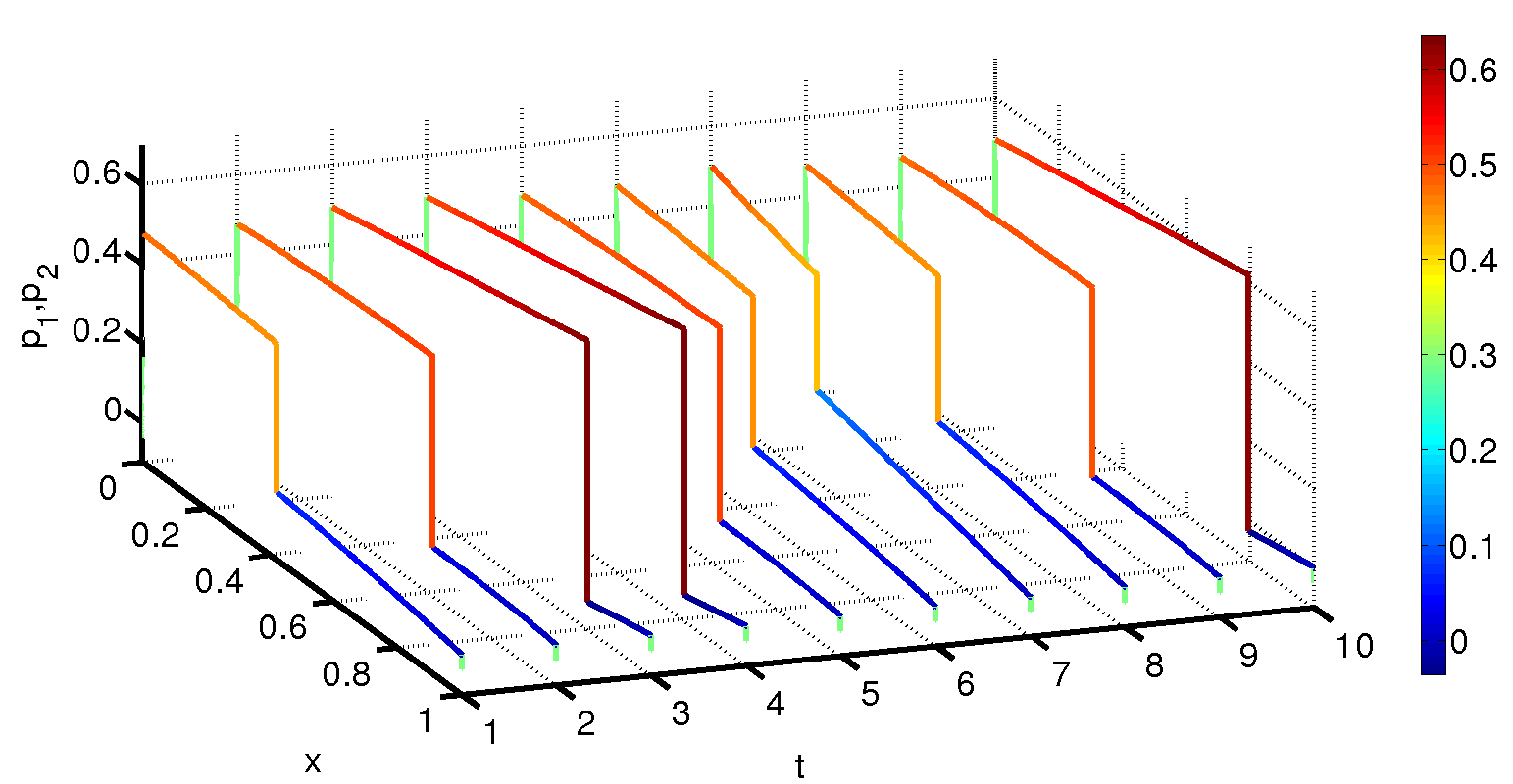}}\\
\subfloat[ \label{fig:tord1sineb}]{\includegraphics[scale=0.25]{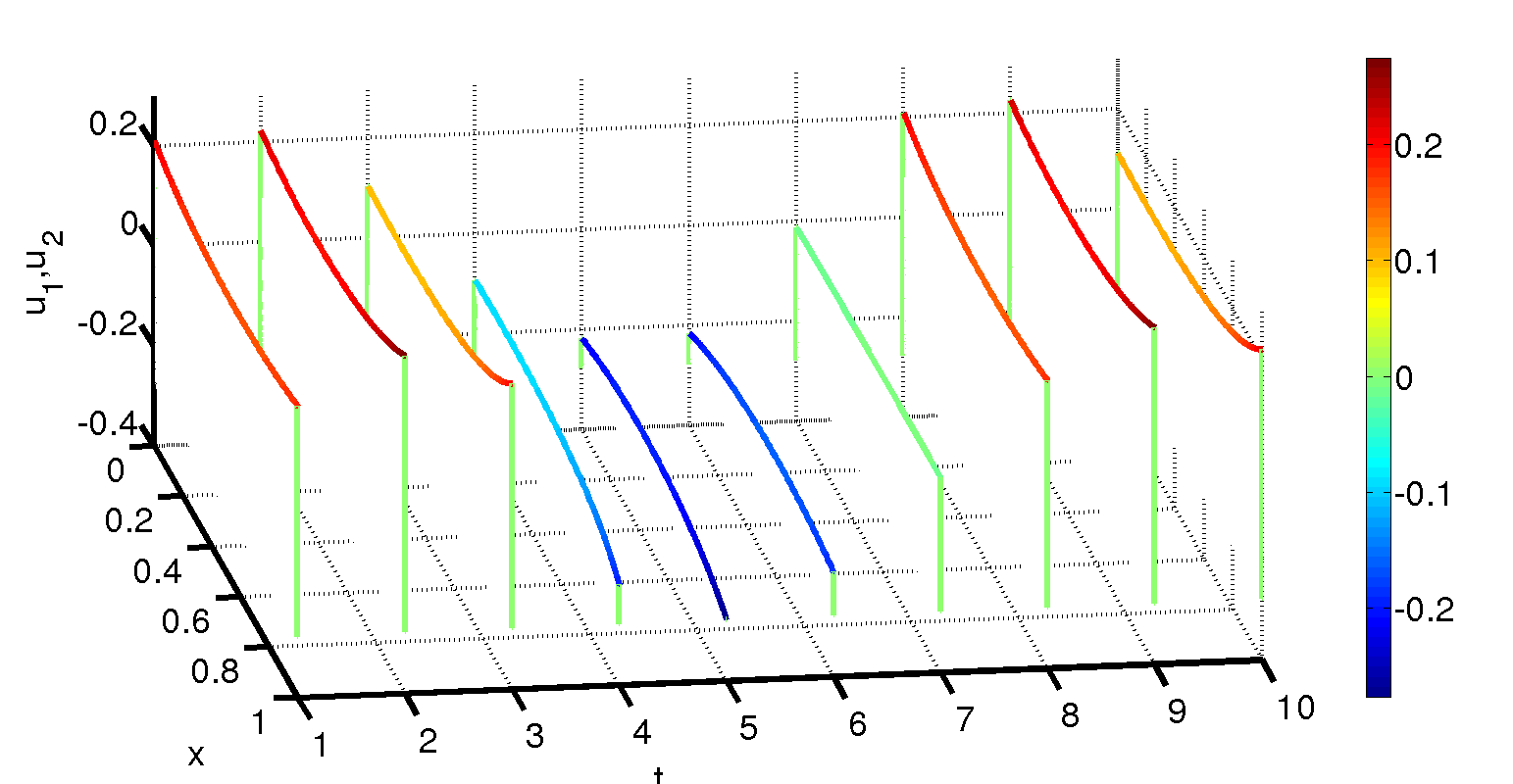}}\\
\subfloat[\label{fig:tord1sinec}]{\includegraphics[scale=0.25]{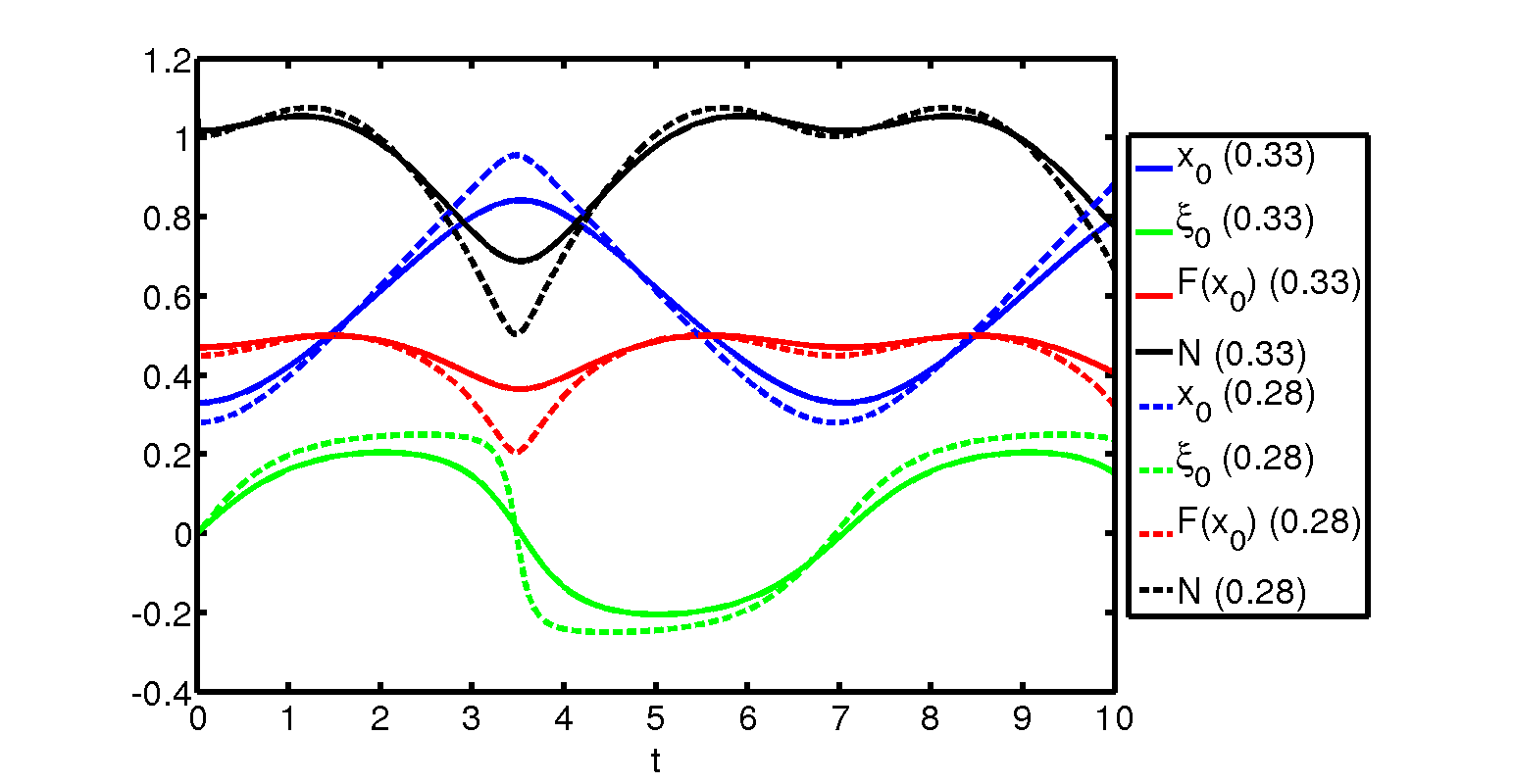}}
\caption{Numerical solutions for the behaviour at $O(1)$ times with fluid effects. Here, the body shape $F$ is sinusoidal, and $M=I=0.125$. (\textit{a}) Pressures $p_1,p_2$ respectively ahead of and behind the contact point for the initial conditions $(x_0,\xi_0)=(0.33,0)$. The discontinuity across the contact point is clearly visible. (\textit{b}) Velocities $u_1,u_2$ respectively ahead of and behind the contact point for the initial conditions $(x_0,\xi_0)=(0.33,0)$. The continuity and relative smoothness at the contact point are noted. (\textit{c}) Contact location $x_0$, its velocity $\xi_0$, its local shape response $F(x_0)$, and the reaction force $N$ versus $t$, for the initial conditions $(x_0,\xi_0)=(0.33,0)$ (solid lines) and $(0.28,0)$ (dashed).}\label{fig:tord1sine}
\end{figure}

\begin{figure}
\centering
\subfloat[\label{fig:tord1ella}]{\includegraphics[scale=0.16]{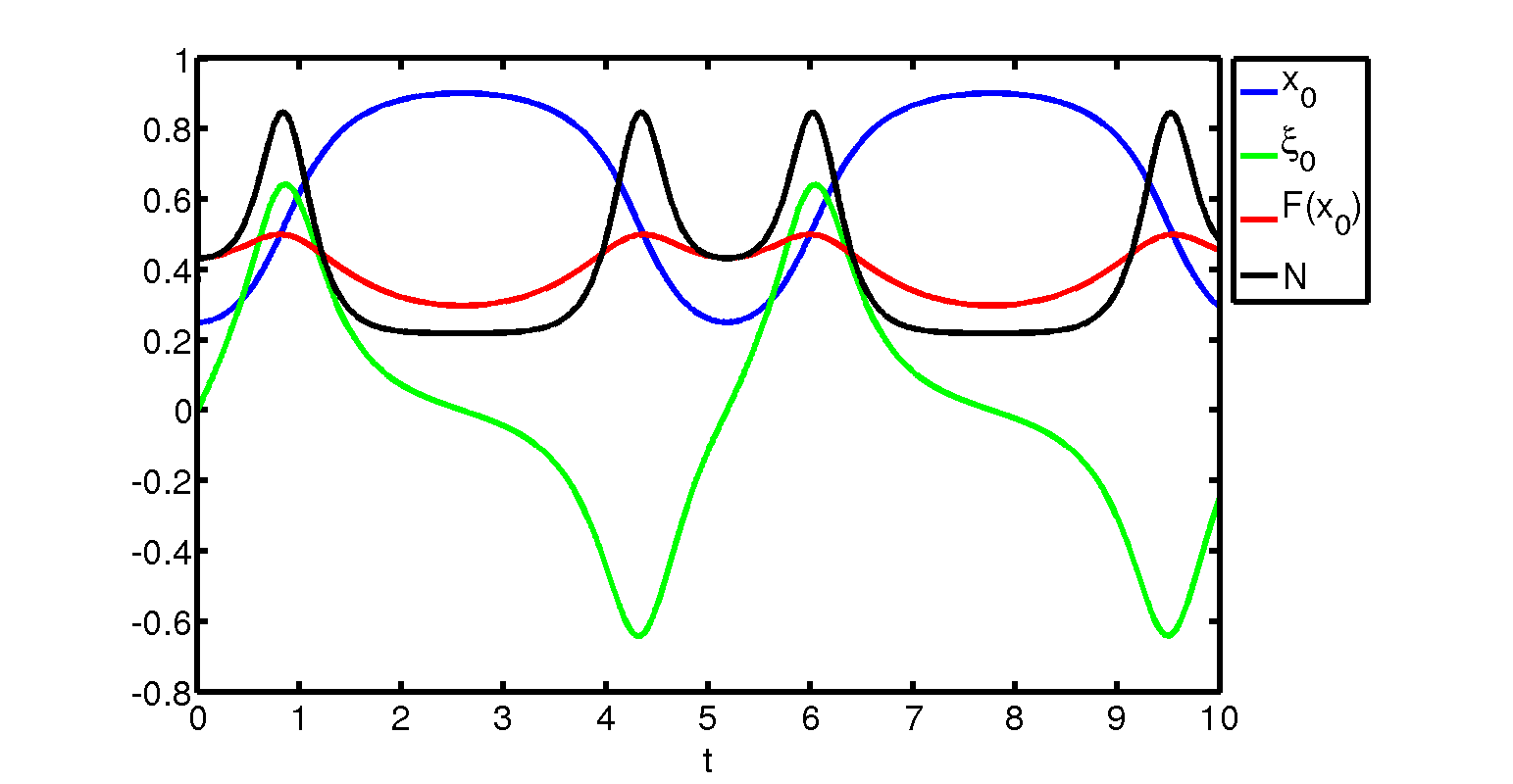}}
\subfloat[\label{fig:tord1ellb}]{\includegraphics[scale=0.16]{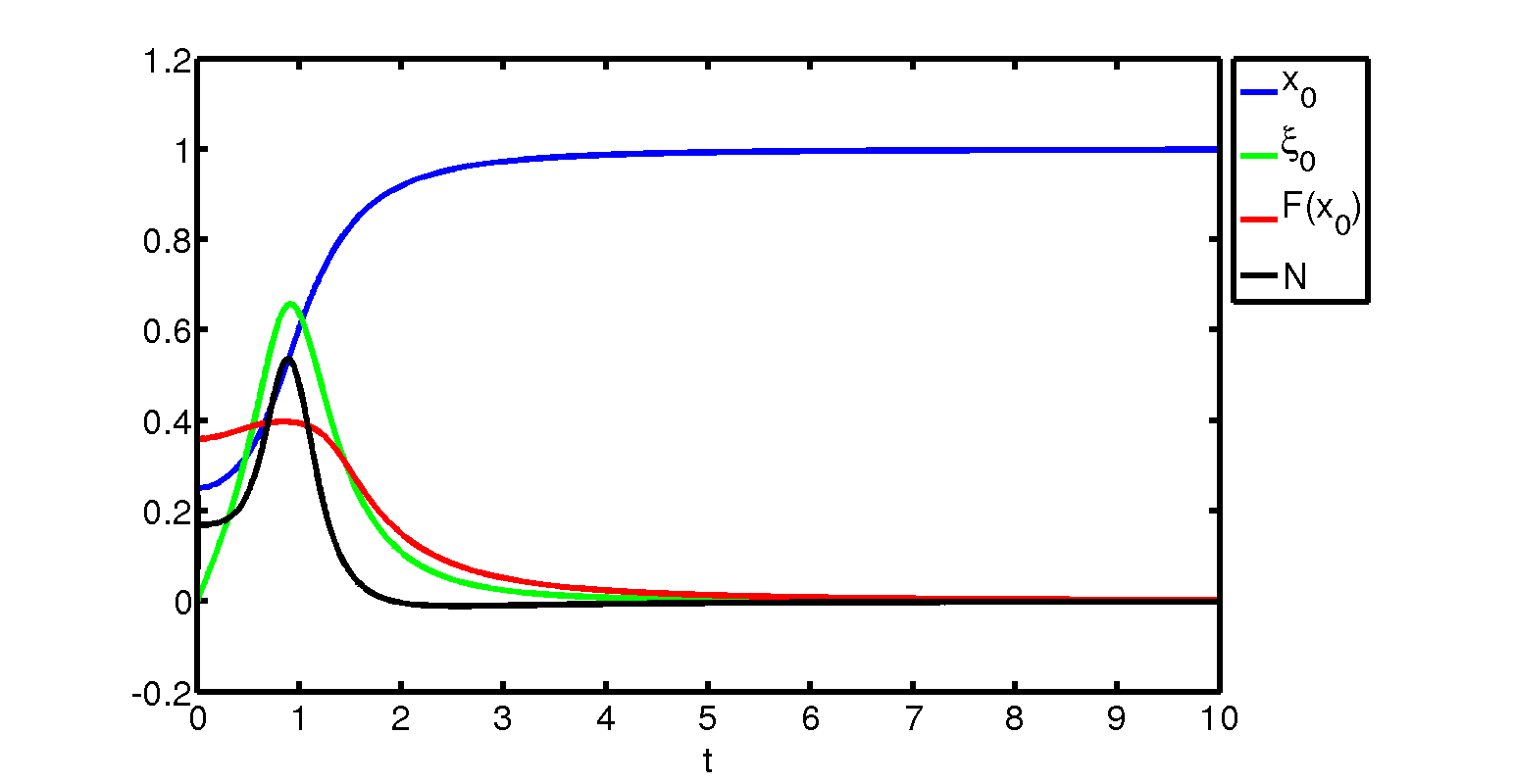}}\\
\subfloat[\label{fig:tord1ellc}]{\includegraphics[scale=0.16]{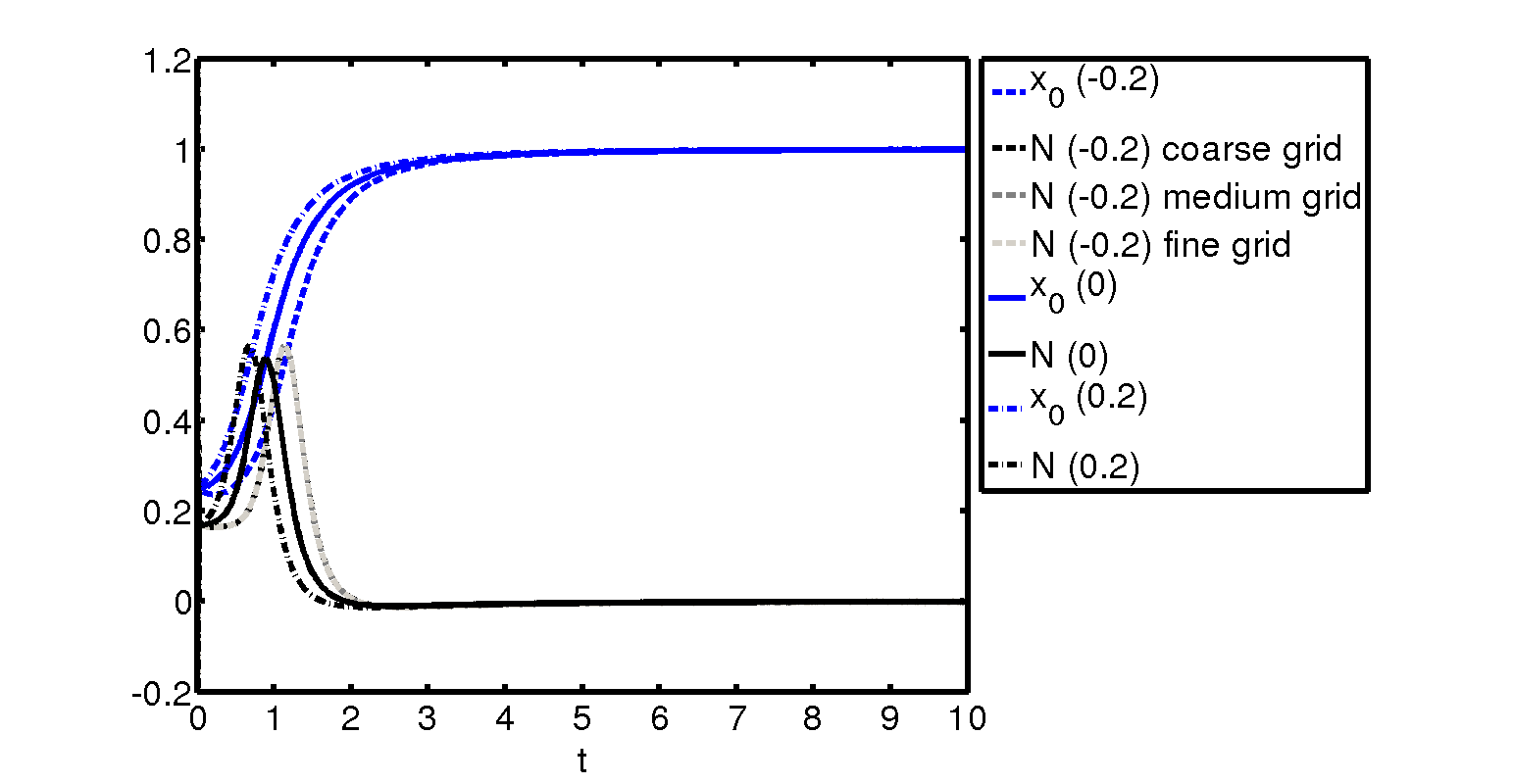}}
\subfloat[\label{fig:tord1elld}]{\includegraphics[scale=0.16]{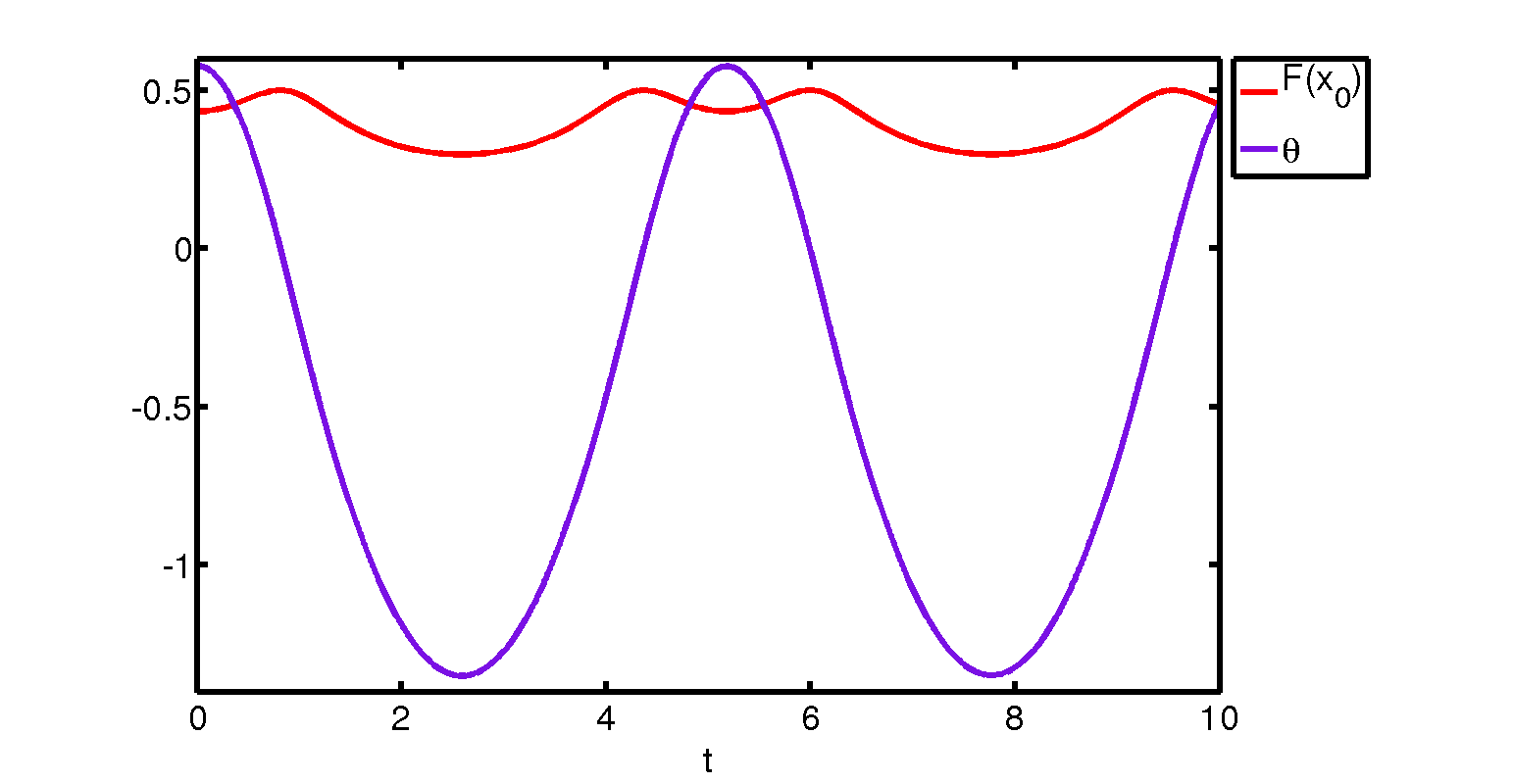}}
\caption{Numerical solutions for the behaviour at $O(1)$ times with fluid effects. Here, the body shape $F$ is elliptical. (\textit{a}) Here, for the initial condition $(x_0,\xi_0)=(0.25,0)$ and for $M=I=0.08$, are shown the contact location, its velocity, its local shape response, and the reaction force plotted against scaled time. (\textit{b}) As for (\textit{a}) but with $M=I=0.05$. (\textit{c}) Results for $x_0,N$ for three different initial contact velocities $\xi_0$ equal to $-0.2,0,0.2$, with the same settings as in (\textit{b}). Responses for the first value calculated with three different time steps are also shown (labelled ``coarse grid'', ``medium grid'', and ``fine grid''). (\textit{d}) The evolution of $F(x_0)$ and $\theta$ with time are shown for the same conditions as in figure \ref{fig:tord1ella}.}\label{fig:tord1ell}
\end{figure}

\begin{figure}
\centering
\includegraphics[scale=0.25]{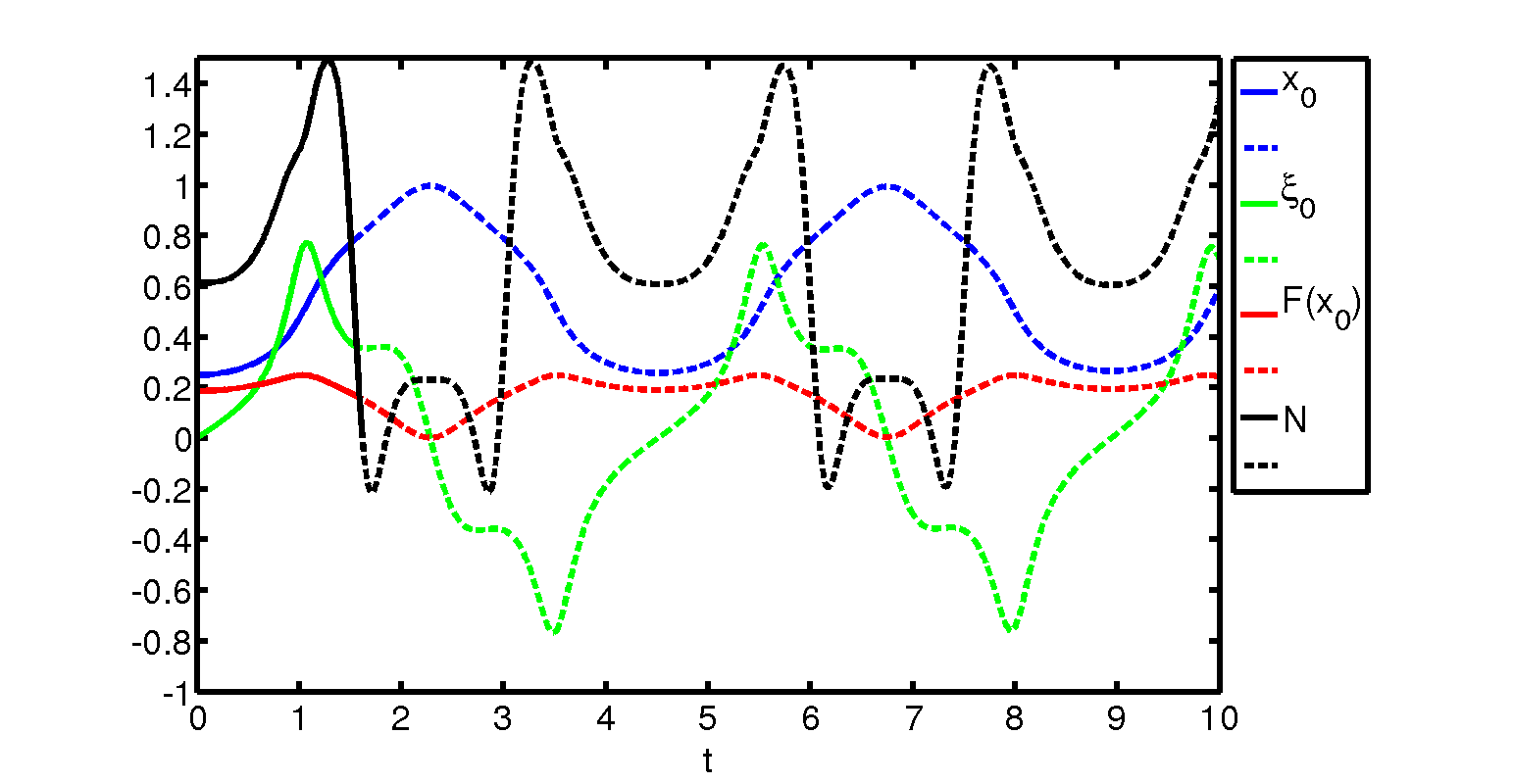}
\caption{Numerical solutions for the behaviour at $O(1)$ times with fluid effects. Here, the body shape $F$ is the smooth-cornered body of figure \ref{fig:smalltc}, with $M=I=0.1$ and initial conditions $(x_0,\xi_0)=(0.25,0)$. Beyond the lift-off time when $N$ first becomes negative, numerical results are shown by dashed lines since strictly they are no longer physically meaningful even though interesting.}\label{fig:tord1smooth}
\end{figure}

Numerical solutions of (\ref{eq:tord1}) were derived by adding the fluid effects of (\ref{eq:ize}) (the added mass, as it were) in lagged style into (\ref{eq:tord1a}), then solving that for an updated $x_0$, feeding this latest $x_0$ into the fluid-flow calculation applied upstream and downstream of the $x_0$-station to determine $u$ from (\ref{eq:gap1h}), (\ref{eq:gap2h}) and hence $p$ from (\ref{eq:gap1u},c),(\ref{eq:gap2u},c), allowing the latest $p$ to be fed into (\ref{eq:ize}), and so on, iterating per time step. An interpolation was employed in the flow calculation to handle the movement of the contact point $x=x_0$ with time, making use of the requirements (\ref{eq:gap1attach}),(\ref{eq:gap2attach}) and enabling the quasi-mass flux to be evaluated. The typical values of the uniform time step $\delta t$ and spatial step $\delta x$ used in the computations were 0.001 and 0.005 in turn, and the effects on the solutions of varying the steps were tested thoroughly. The major results are presented in figures \ref{fig:tord1sine}--\ref{fig:tord1smooth} for various prescribed shapes $F$ and varied other conditions.

Figure \ref{fig:tord1sine} is for the sinusoidal shape, and $M=I=0.125$. Figures \ref{fig:tord1sine}a,b, in which the initial conditions are of $x_0$ being 0.33 while the contact velocity $\xi_0$ is zero, show details of the evolving pressures and velocities in the fluid-filled gaps over a scaled time range of 0--10. Thus figure \ref{fig:tord1sinea} has the results for pressures $p_1,p_2$ ahead of and behind the contact point plotted against $x$ at integer times, and figure \ref{fig:tord1sineb} gives the respective fluid velocities $u_1,u_2$ versus $x$ . It is interesting that the pressure curves upstream and downstream of contact are almost straight with only a slight curvature being apparent. Also the vertical lines in figure \ref{fig:tord1sinea} mark the clear jumps in pressure at the moving contact position, in keeping with the earlier remarks. The fluid velocities $u_1,u_2$ in contrast seen in figure \ref{fig:tord1sineb} are continuous through the contact position $x=x_0(t)$ as anticipated previously. Further, the velocities are all positive at early times but eventually negative values are encountered (smoothly) at later times as the front of the body rocks downwards squashing the fluid there in a sense; the inferred flow closer to the leading edge is then clockwise around the leading edge. Figure \ref{fig:tord1sinec} presents the numerical results for the evolution of the contact location $x_0$, its velocity $\xi_0$, its local shape response $F(x_0)$, and the reaction force $N$ versus $t$, both for the initial condition $(x_0,\xi_0)$ of $(0.33,0)$ as above, and for $(0.28,0)$ for comparison. The latter case shows a phenomenon of rapid rocking appearing near the end points which is not dissimilar to that encountered in no-fluid scenarios subject to different initial conditions.

In figure \ref{fig:tord1ell} the results for the elliptical shape are presented. Figure \ref{fig:tord1ella} with an initial condition $(x_0,\xi_0)$ of $(0.25,0)$ is for $M=I=0.08$ and shows the contact location, its velocity, its local shape response, and the reaction force plotted against scaled time. The findings are fairly similar to those for the no-fluid case in figure \ref{fig:tord1noflow1} but clearly the fluid effect is no longer negligible here, although an apparently periodic rocking behaviour emerges nonetheless. Figure \ref{fig:tord1ellb} has the same initial condition but now $M=I=0.05$, implying increased fluid-flow effect. Here the contact position $x_0$ is found to increase monotonically with time, and tends to approach the trailing edge, but during that process the reaction force $N$ becomes negative and so lift-off of the body is indicated. This arises at a finite time $t$ which in the case of figure \ref{fig:tord1ellb} is at $t=1.921$ and the final contact at that time is at an $x_0$ value of 0.9096. Figure \ref{fig:tord1ellc} reinforces the point and further examines the effects of the initial conditions by showing the results for $x_0,N$ for three different initial contact velocities $\xi_0$ equal to $-0.2,0,0.2$, with the same settings as in \ref{fig:tord1ellb} and with the responses for the first value also being calculated with three different time steps to highlight the accuracy involved. In every case the reaction force becomes negative within a finite scaled time. The evolution of the angle $\theta$ under the conditions of figure \ref{fig:tord1ella} is presented in figure \ref{fig:tord1elld} and also supports the conclusions on nonlinear periodic rocking behaviour if lift-off is absent, as well as agreeing with the small-time analysis of section \ref{sec:smallt}. In figure \ref{fig:tord1elld} the successive maxima in $F$ versus time are seen to correspond to $\theta$ changing sign, whereas the minima in $F$ correspond to successive extrema in $\theta$ because of the relationship between $F$ and the contact location $x_0$.

The smooth-cornered shape of figure \ref{fig:smalltc} is the subject of figure \ref{fig:tord1smooth} with fluid-flow influences now being present. The trend observed is akin to that described in the previous paragraph, including the monotonic response of the contact location prior to the eventual lift-off inferred from the reaction force $N$ becoming negative at a finite time. Numerical results generated from the lift-off time are shown by dashed lines to indicate that they are not strictly physically meaningful even though interesting.

\begin{figure}
\centering
\includegraphics[scale=0.25]{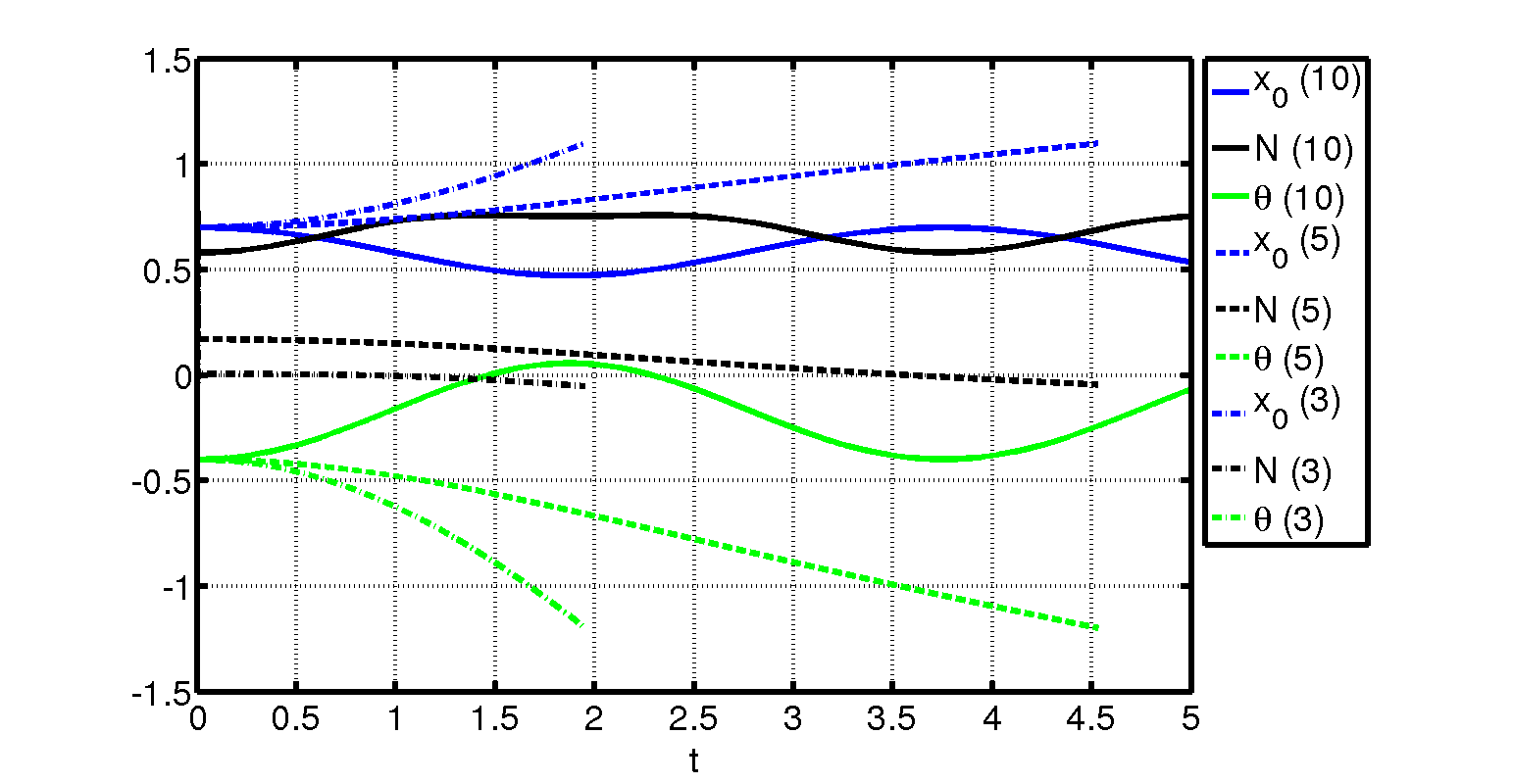}
\caption{The influence of the gravity parameter $g^+$ (value in parentheses in legend) on the behaviour of the constant-curvature body of figures \ref{fig:smalltd}--\ref{fig:smalltf}. Here, $M=I=0.1$, and the initial conditions are $(x_0,\xi_0)=(0.7,0)$. }\label{fig:g}
\end{figure}

The influence of the parameter $g^+$ was examined first in figures \ref{fig:smallte} and \ref{fig:smalltf} and is re-examined in figure \ref{fig:g}. Reducing $g^+$ from its usual value of 10 to the values 5 and then 3 is found to result in a change from periodic rocking behaviour to finite-time lift-off as $N$ becomes negative as in figure \ref{fig:g}, doing so earlier for the value 3 (at $t=0.817$) than for the value 5 (at $t=3.576$). The contact position moves monotonically towards the trailing edge for the two reduced $g^+$ values, and the associated $\theta$ angles decrease monotonically, until lift-off is encountered. The lift-off is perhaps the most intriguing phenomenon to explore next.

\section{Lift-off of the body}\label{sec:liftoff}

The evidence from the results in figures \ref{fig:tord1ell}--\ref{fig:g} indicates that the interactive solution is regular at the onset of lift-off of the body, with (to repeat) lift-off being identified with the reaction force becoming zero after being positive beforehand. Thus the expression for the flow and body-movement behaviour at the lift-off time $t_\text{LO}$ say has

\begin{equation}\label{eq:lomode}
(h, \theta, u, p, x_0, N) = \sum_{m=0}^\infty (h, \theta, u, p, x_0, N)^{(m)} (t_\text{LO} - t)^m\ .
\end{equation}

The coefficients $h^{(m)}$ etc on the right-hand side for $m \geqslant 0$ are independent of $t$ but their values depend on the evolution from the initial time, while (\ref{eq:lomode}) holds for scaled time $t$ tending to $t_\text{LO}-$ , the lift-off time being positive. It is noticed from figure \ref{fig:config} and so on that the flow velocity $u$ and pressure $p$ are in two parts I, II on either side of the contact point. For the pressure in particular this facet implies there is a moving discontinuity to be dealt with as highlighted by figure \ref{fig:tord1sinea} and hence a little delicacy near the terminal location $x = x_0(t_\text{LO})$ at small relative times $t_\text{LO}- t$, but this does not affect the response (\ref{eq:lomode}) materially. 

Substitution of (\ref{eq:lomode}) into the fluid-body equations of section \ref{sec:fbi} or into the equivalent system (\ref{eq:tord1a},b) can then be made and studied in detail. The principal point however is that in (\ref{eq:tord1a},b) the integrated flow-pressure contributions $i_1$, $i_2$ act to counter the effect of scaled gravity on the right-hand side and thus take the interactive solution into a part of solution space which is not attainable in the no-fluid scenario of section \ref{sec:ord1tnoflow}. The action which is due to added mass and to evolution in the fluid-body interaction allows the reaction force to become negative. The same global type of anti-gravity action is also responsible for the changes of sign in the reaction force at early times (section \ref{sec:smallt}) as shown in the results of figure \ref{fig:smallt}.

The results for the ellipse in (\ref{eq:limitellipse}), figure \ref{fig:tord1noflowb} without fluid effects and subsequently figures \ref{fig:tord1ell}(a-c) with fluid effects provide a specific example of the above. Checking on the orders of magnitude of gap properties and flow behaviour locally near the contact point whether close to the leading edge ($x = \epsilon X$) or to the trailing edge shows $dh/dt$ and $d\theta/dt$ to be of typical size $ \epsilon^{-\frac{1}{4}}$ which is large. So $\Gamma$ is of the same typical size. The fluid velocity $u$ is then of order $dx_0/dt$, that is, $ \epsilon^{\frac{5}{4}}$, making $q$ also small of order $ \epsilon^{\frac{5}{2}}$. The local pressure response $p$ is therefore merely a small perturbation from the value $1/2$ and hence has negligible influence on $i_1$, $i_2$ in (\ref{eq:tord1a},b)  as these remain overwhelmed by global contributions of order unity. Similar reasoning applies for the sinusoidal shape of (\ref{eq:limitsine}) and figures \ref{fig:tord1noflowa}, \ref{fig:tord1sine}(a-c) for example.

The conclusion then is there is nothing dramatic about the onset of lift-off, in line with (\ref{eq:lomode}). In contrast such a conclusion is not necessarily so once lift-off occurs for times $t$ equal to $t_\text{LO}+$: the latter is considered within the next section. As regards the work so far the more significant question is whether lift-off occurs or not and this is clearly reliant on the interaction parameters together with initial values. According to the model lift-off generally cannot take place without fluid flow effects but it can with them.

\section{Comparisons and final comments}\label{sec:conc}

The lift-off of a body from a fixed solid surface due to fluid motion or indeed just the washing or rocking of such a surface-mounted body clearly depends on a significant number of parameters.  The total parameter space should also include the influences of evolution (the initial-value problem due to wash starting up or wind changing for example) and body shape because of their importance. The present investigation it is hoped helps to shed light on this parameter space by means of the specific studies in sections \ref{sec:smallt}--\ref{sec:tord1} on small-time responses, zero-fluid evolution and with-fluid evolution in turn, given the model set up in section \ref{sec:fbi} and the account of lift-off in section \ref{sec:liftoff}.  In addition concerning scaled gravity effects in particular at low Richardson number two modes of lift-off can now be identified. One occurs for enhanced values of the scaled mass and so has the gravitational force in balance with the mass-acceleration contribution on account of a reduction in the local variation in gap width. Typically this mode corresponds in dimensional terms to increased incident fluid speed (including threshold speed) or increased body mass. The second mode occurs for in effect zero scaled gravity in the sense that the gravitational force simply has negligible influence on the mass-acceleration balance compared with the flow forces and the normal reaction force. Such a second mode can be associated mainly with relatively reduced body or particle dimensions or negligible gravity in reality. The specific exact case addressed in appendix \ref{app:froude} backs up the presence of these two modes of lift-off. 

It is interesting to return now to consider the movement of sand or dust on the surface of Mars, a matter introduced in section \ref{sec:intro}. As a broad background observations have suggested that the flux of Martian dust or sand movement is comparable overall to that on Earth.  There are, obviously or potentially, many physical factors at work here as is indicated also in the substantial fairly recent growth in studies: again see section \ref{sec:intro}. The present study suggests in fact there are wide areas of parameter space where lift-off can occur in practice. 

Two attributes which might be emphasised first are that the Martian gravity $g_D$ is about $0.38$ of the Earth value whereas the density of the Martian atmosphere $\rho_D$ is only about $1/60$ of Earth value. Predictions then from our results would be based on knowing that $M = M_D \bar{h}/(\rho_D L_D^2)$ and $g^+ = g_D L_D/(\bar{h} U_D^2)$ from the non-dimensionalisation, and estimating the body mass as $M_D = \rho_{D\text{body}} L_D^2\times\text{(unit distance in }z_D)$, where $\rho_{D\text{body}}$ is the typical body density. In consequence
\begin{equation}\label{eq:marsmass}
M = \frac{\rho_{D\text{body}}}{\rho_D} \bar{h}^2\ .
\end{equation}
Thus on Mars, taking the same size, shape and density of body (dust particle) as on Earth, one should tend to find $M$ increasing by virtue of the $\rho_D$ factor in $M$ although readily mitigated by the gap width $\bar{h}$ reducing by a factor $1/4$ say, making $M$ increase by about 4. Now consider the mode mentioned earlier for lift-off assuming $M$ is relatively large, that is, $M \sim 1/g^+$ or $M < 1 /g^+$. This becomes the requirement $U_D^2 > (\rho_{D\text{body}} / \rho_D) g_D \bar{h} L_D$ on the threshold wind speed $U_D$ for lift-off to occur. In numerical terms the right-hand side here as far as Mars is concerned is approximately $(60) \times 0.4 \times 1/4 \times 1$ relative to the Earth value in view of the estimates above, i.e. the Mars value is about 6 times the Earth value on the right-hand side. Hence the critical wind speed on Mars is predicted as about 2--3 times that on Earth. This last result is reasonably in line with the references, e.g.  \citet{wang12}; further the current prediction of a square-root dependence of threshold speed on particle size is not out of keeping with figure 2 of that last paper.

Second, though, is the need for caution, since the above prediction and indeed earlier predictions should be qualified heavily due to the following aspects. As we have shown, lift-off can appear quite readily at low scaled mass or large scaled mass typically: see also figure \ref{fig:smallte}. Moreover shape effects, initial conditions and gap width all matter considerably as shown in the current study. Many areas of the total parameter space allow lift-off to occur. Again in reality the `typical' body shapes vary appreciably rather than being purely spherical or thin, while with many bodies or particles present there is a potentially complicated and subtle multi-body factor where for example a string of bodies lies on top of another string, creating a thin gap, and we should not forget the existence of a laminar or turbulent boundary layer in the incident fluid motion. Finally here, as shown in the previous section, lift-off itself is not a dramatic event in terms of the fluid-body interaction per se: it merely happens in a smooth manner, at least in the pre-lift-off stage (compare with the post-lift-off stage addressed below). This manner is quite in keeping with the stated importance of the initial conditions. Parameter studies of reptation in terms of fluid-body interaction along lines similar to those here or in \citet{smithwilson11} could be of much interest.

Intriguing issues remain. The influence of incident shear for example in the depths of an oncoming atmospheric boundary layer has still to be considered. A classic treatment of this aspect is for a uniform shear. In addition the modelling so far has ignored viscous effects quite generally. Other shapes or configurations of body need to be examined such as cases with non-symmetry in the streamwise direction or with irregular shapes or groups of bodies, in order to raise understanding for dust piles for instance. (Appendix \ref{app:froude} shows that the body curvature can act in effect to increase the scaled mass parameter and reduce the gravity force, which leads on again to another part of parameter space.)

Properties arising after a lift-off are also intriguing. In cases where lift-off does occur (negating the energy-integral result of the no-flow case) the local scales soon after lift-off are expected to be similar to those in the current study, such as a relative time squared scaling in the vertical coordinate. This would suggest that a local analysis then has to deal with a substantial jump in pressure within which the local pressure is an order-one function of distance measured from the lift-off point but scaled with relative time. In some detail near a moving contact point prior to lift-off of a smooth body the local gap shape is typically of the form $H \sim \mu(t) \bar{x}^2$ if the contact point is at $x = x_0(t)$. Here $\mu(t)$ is an $O(1)$ function of time which is arbitrary in the sense of being determined by historical effects, and $\bar{x}$ denotes $x- x_0(t)$. Now the response in horizontal fluid velocity is of the form $u \sim x_0' +  r \bar{x}$ where the rate $r(t)$ is $\mu'/(3 \mu)$, implying that the pressure behaves as $c_2(t) + x_0'' \bar{x} + (r'+r^2) \bar{x}^2/2$  in the vicinity of the contact point. However the $c_2$ pressure function is discontinuous across the contact position, as is clear from the results in figure \ref{fig:tord1sinea} for example and the working in earlier sections. This reinforces the suggestion of a pressure discontinuity which, immediately after lift-off, must be accommodated by the fluid flow through the newly opened gap.

The study has been focussed throughout on unsteady nonlinear interactions in just two spatial dimensions. Moving the theory on to three-dimensional interactions is called for. The three-dimensional version of the post-lift-off configuration described in the previous paragraph for example would be of much interest. Many other extensions naturally suggest themselves: allowing body flexibility (with applications to red blood cell deformation and swimming microorganisms amongst others); studying the effects of surface shape, curvature, and roughness; and including the effects of a thin layer of a second, viscous fluid between the surface and rocking body, to study the effects of surface tension there.

Thanks are due to members of the Medical Modelling Group at UCL and the UK Icing Group for their interest and comments, and to Dr Miguel A. Moyers Gonzalez at UC for useful discussions.

\appendix

\section{The constant-curvature body at small times}\label{app:froude}

\begin{figure}
\centering
\subfloat[\label{fig:Aa}]{\includegraphics[scale=0.25]{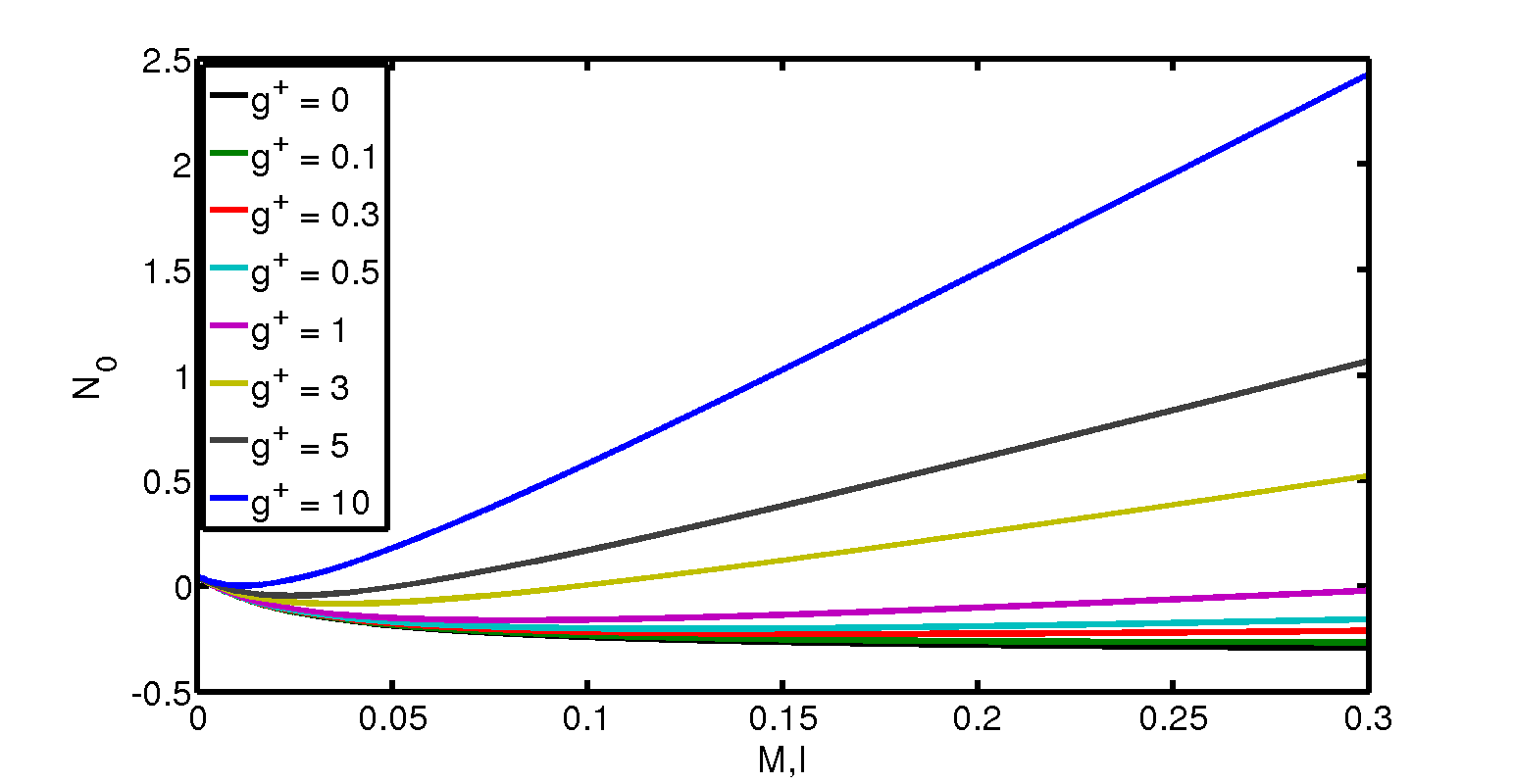}}\\
\subfloat[\label{fig:Ab}]{\includegraphics[scale=0.25]{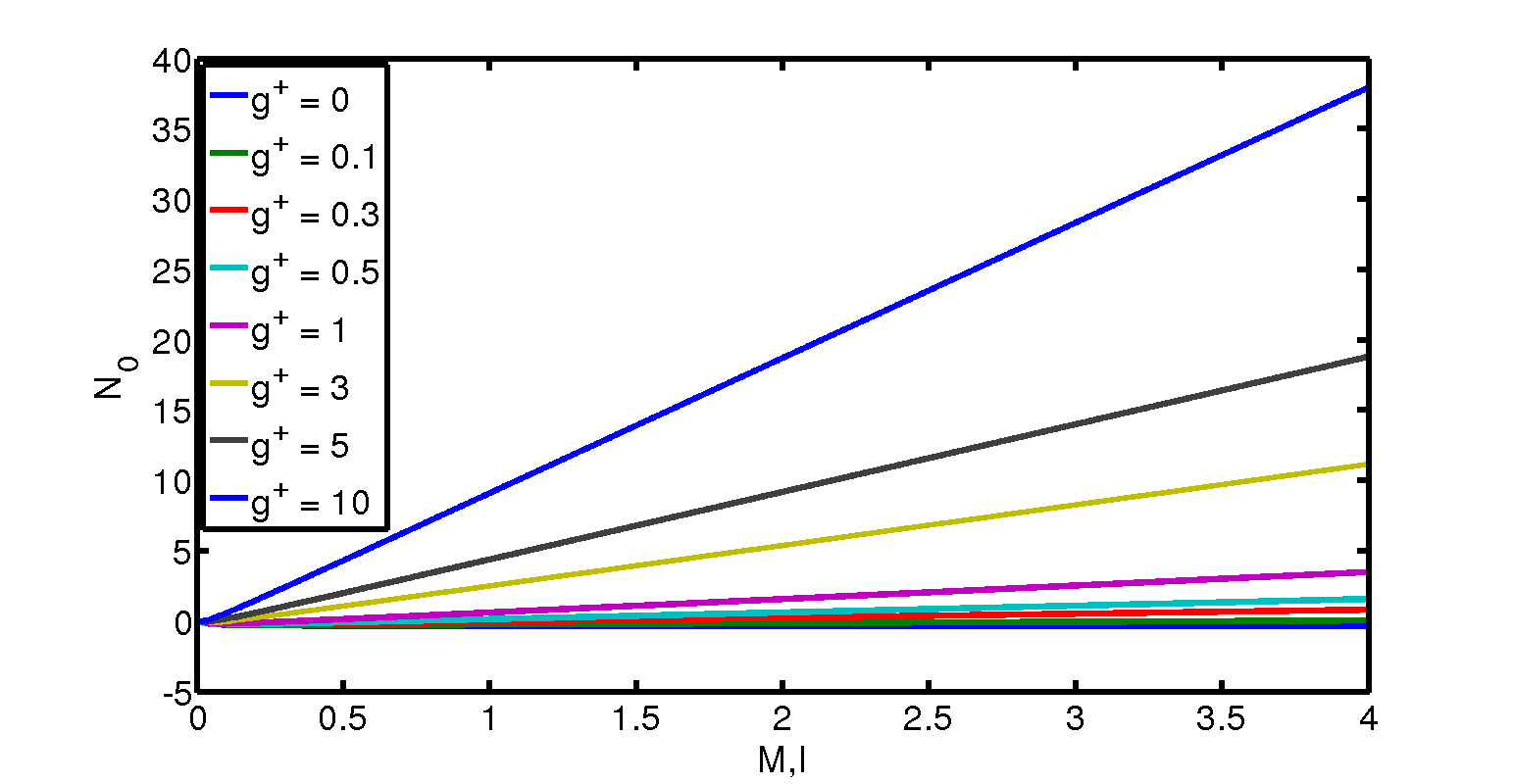}}\\
\subfloat[\label{fig:Ac}]{\includegraphics[scale=0.25]{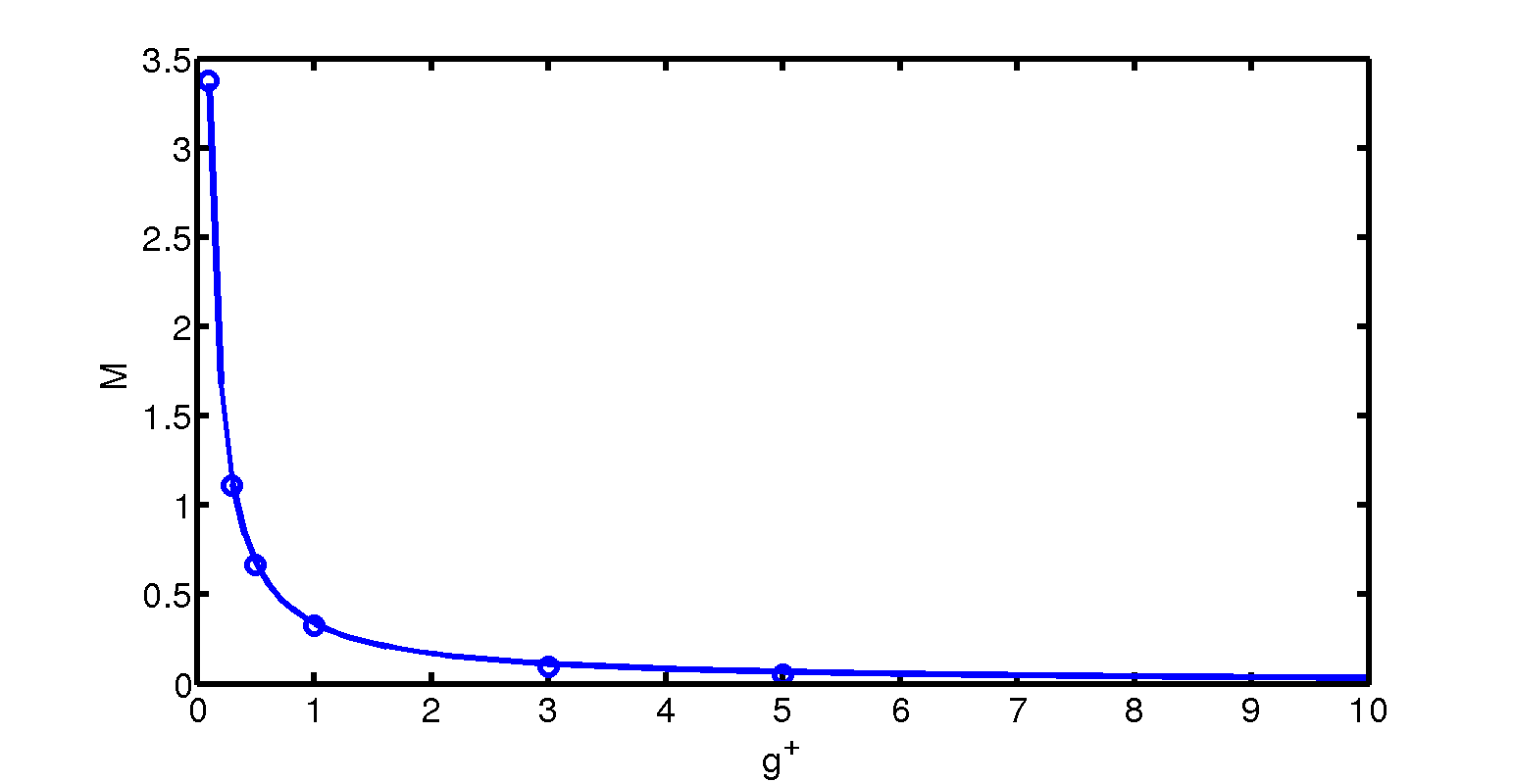}}
\caption{The constant curvature body at small times. (\textit{a}) Solutions of equation (\ref{eq:n0explicit}) for a range of values of $g^+$. (\textit{b}) As for figure \ref{fig:Aa} but for a wider range of values of $M=I$. (\textit{c}) The numerical roots of (\ref{eq:n0explicit}) obtained from figures \ref{fig:Aa}, \ref{fig:Ab} (circles) compared with the asymptotic prediction of equation (\ref{eq:masym}) (solid curve). }\label{fig:A}
\end{figure}

A constant curvature in the present setting corresponds to having the initial gap width
\begin{equation}
H_0(x) = \lambda (x-A)^2\ ,
\end{equation}
where $\lambda$ is a positive constant proportional to the body curvature. Hence evaluating the integrals in (\ref{eq:p0eq})--(\ref{eq:r}) and in particular finding that $r$ is $(-A^2 + A - 1/3) /(2 \lambda)$ leads to the result
\begin{equation}\label{eq:apps}
\left(\frac{A^2  - A + \frac{1}{3}}{2\lambda} + 2 I + 2M(A -\frac{1}{2})^2\right) \theta_2 = s         
\end{equation}
for $\theta_2$ in terms of $s$ which is $Mg^+(A-1/2) - A^2/4$. The value of $x_c$ is taken as $1/2$ as a major example. Then the scaled reaction force $N_0$ follows from (\ref{eq:noI}) in the form
\begin{equation}
N_0 = \frac{2 I \theta_2 - J_2}{A -\frac{1}{2}}\ .  
\end{equation}
Here $J_2$ can be determined analytically from (\ref{eq:j2}) with (\ref{eq:p0eq}), (\ref{eq:P0eq}) and so this yields the explicit solution
\begin{equation}\label{eq:n0explicit}
N_0 = \frac{\left(2 I - \frac{A^2- A + \frac{1}{6}}{2\lambda} \right)\theta_2  -\frac{A^2 -A}{4}}{A -\frac{1}{2}}
\end{equation}
for $N_0$, with $\theta_2$ given explicitly by (\ref{eq:apps}).  It can be shown by working with the quantities $\lambda M$, $\theta_2/\lambda$, $g^+/\lambda$ that $\lambda$ may be normalised to unity without loss of generality.

Plots of the scaled force $N_0$ against the scaled mass $M$ are presented in figures \ref{fig:Aa} and \ref{fig:Ab} for varying values of the gravity contribution $g^+$.  In these plots the scaled moment of inertia $I$ is again equated to $M$ for reasons described earlier on and the curvature constant $\lambda$ is kept at unity while the starting location $A$ is kept at 0.7 for comparison purposes. The results here agree with those presented in section \ref{sec:smallt}.

The lift-off case where $N_0$ is zero is of much interest of course. This critical case is controlled by the relation 
\begin{equation}\label{eq:critrel}
2 \mu g^+ M^2 - \left[\mu A + \frac{A(A-1)(A-\frac{1}{2})}{2} + \frac{(A^2 - A + \frac{1}{6}) g^+}{2\lambda}\right] M + \frac{(A -\frac{2}{3})A}{8\lambda} = 0 
\end{equation}
between $M$, $g^+$, $A$ and $\lambda$, from (\ref{eq:n0explicit}) with (\ref{eq:apps}), with $\mu$ standing for the ratio $I/M$. When the gravity effect is comparatively small for instance the first two contributions are overwhelming when $M$ is large and the relation (\ref{eq:critrel}) leads to the explicit form
\begin{equation}\label{eq:masym}
M \sim \frac{[2\mu  + (A-1)(A -\frac{1}{2})] A}{4 \mu g^+}\qquad   \text{(first root for $g^+$ small)}
\end{equation}
for the dependence of the critical $M$ value on $g^+$ in particular. The numerator here can be positive or negative, depending on the ratio $\mu$ and the contact position $A$, which indicates a sensitivity to the precise body shape. However for the values $\mu=1$ and $A = 0.7$ used in figures \ref{fig:Aa} and \ref{fig:Ab} the asymptotic trend (\ref{eq:masym}) is found to agree rather closely with the full numerical results in the figures as $g^+$ varies and indeed the asymptote even works well for $g^+$ equal to unity or greater, as indicated by figure \ref{fig:Ac} . The trend (\ref{eq:masym}) for this special category of body shapes also hints at the trend for the general body shape, which again is found to have $M$ proportional to the inverse of effective gravity.

Similar considerations apply to the other root for relatively small gravity in (\ref{eq:critrel}), giving the explicit value
\begin{equation}
M \sim \frac{(A - \frac{2}{3})A}{8\lambda[\mu A + \frac{A(A-1)(A-\frac{1}{2})}{2}]}\qquad  \text{(second root for for $g^+$ small).}
\end{equation}
This is associated with having identically zero gravity in (\ref{eq:critrel}) and corresponds to one of the cases presented in figure \ref{fig:A}. The potential reliance on $A$ exceeding 2/3 is again noted.

\bibliographystyle{plainnat}

\bibliography{grain}

\end{document}